\documentclass[ALICE,manyauthors]{cernphprep}
\usepackage[comma,square,numbers,sort&compress]{natbib}
\usepackage{hyperref}
\usepackage{lineno}
\usepackage{xspace}
\usepackage[dvipsnames]{xcolor}
\usepackage{amsmath}
\usepackage{upgreek}
\usepackage{orcidlink}

\begin{document}
%

\newcommand{\pp}           {pp\xspace}
\newcommand{\ppbar}        {\mbox{$\mathrm {p\overline{p}}$}\xspace}
\newcommand{\XeXe}         {\mbox{Xe--Xe}\xspace}
\newcommand{\PbPb}         {\mbox{Pb--Pb}\xspace}
\newcommand{\pA}           {\mbox{pA}\xspace}
\newcommand{\pPb}          {\mbox{p--Pb}\xspace}
\newcommand{\AuAu}         {\mbox{Au--Au}\xspace}
\newcommand{\dAu}          {\mbox{d--Au}\xspace}

\newcommand{\s}            {\ensuremath{\sqrt{s}}\xspace}
\newcommand{\snn}          {\ensuremath{\sqrt{s_{\mathrm{NN}}}}\xspace}
\newcommand{\pt}           {\ensuremath{p_{\rm T}}\xspace}
\newcommand{\meanpt}       {$\langle p_{\mathrm{T}}\rangle$\xspace}
\newcommand{\ycms}         {\ensuremath{y_{\rm CMS}}\xspace}
\newcommand{\ylab}         {\ensuremath{y_{\rm lab}}\xspace}
\newcommand{\etarange}[1]  {\mbox{$\left | \eta \right |~<~#1$}}
\newcommand{\yrange}[1]    {\mbox{$\left | y \right |~<~#1$}}
\newcommand{\dndy}         {\ensuremath{\mathrm{d}N_\mathrm{ch}/\mathrm{d}y}\xspace}
\newcommand{\dndeta}       {\ensuremath{\mathrm{d}N_\mathrm{ch}/\mathrm{d}\eta}\xspace}
\newcommand{\avdndeta}     {\ensuremath{\langle\dndeta\rangle}\xspace}
\newcommand{\dNdy}         {\ensuremath{\mathrm{d}N_\mathrm{ch}/\mathrm{d}y}\xspace}
\newcommand{\Npart}        {\ensuremath{N_\mathrm{part}}\xspace}
\newcommand{\Ncoll}        {\ensuremath{N_\mathrm{coll}}\xspace}
\newcommand{\dEdx}         {\ensuremath{\textrm{d}E/\textrm{d}x}\xspace}
\newcommand{\RpPb}         {\ensuremath{R_{\rm pPb}}\xspace}

\newcommand{\nineH}        {$\sqrt{s}~=~0.9$~Te\kern-.1emV\xspace}
\newcommand{\seven}        {$\sqrt{s}~=~7$~Te\kern-.1emV\xspace}
\newcommand{\twoH}         {$\sqrt{s}~=~0.2$~Te\kern-.1emV\xspace}
\newcommand{\twosevensix}  {$\sqrt{s}~=~2.76$~Te\kern-.1emV\xspace}
\newcommand{\five}         {$\sqrt{s}~=~5.02$~Te\kern-.1emV\xspace}
\newcommand{\twosevensixnn}{$\sqrt{s_{\mathrm{NN}}}~=~2.76$~Te\kern-.1emV\xspace}
\newcommand{\fivenn}       {$\sqrt{s_{\mathrm{NN}}}~=~5.02$~Te\kern-.1emV\xspace}
\newcommand{\LT}           {L{\'e}vy-Tsallis\xspace}
\newcommand{\GeVc}         {Ge\kern-.1emV/$c$\xspace}
\newcommand{\MeVc}         {Me\kern-.1emV/$c$\xspace}
\newcommand{\TeV}          {Te\kern-.1emV\xspace}
\newcommand{\GeV}          {Ge\kern-.1emV\xspace}
\newcommand{\MeV}          {Me\kern-.1emV\xspace}
\newcommand{\GeVmass}      {Ge\kern-.2emV/$c^2$\xspace}
\newcommand{\MeVmass}      {Me\kern-.2emV/$c^2$\xspace}
\newcommand{\lumi}         {\ensuremath{\mathcal{L}}\xspace}

\newcommand{\ITS}          {\rm{ITS}\xspace}
\newcommand{\TOF}          {\rm{TOF}\xspace}
\newcommand{\ZDC}          {\rm{ZDC}\xspace}
\newcommand{\ZDCs}         {\rm{ZDCs}\xspace}
\newcommand{\ZNA}          {\rm{ZNA}\xspace}
\newcommand{\ZNC}          {\rm{ZNC}\xspace}
\newcommand{\SPD}          {\rm{SPD}\xspace}
\newcommand{\SDD}          {\rm{SDD}\xspace}
\newcommand{\SSD}          {\rm{SSD}\xspace}
\newcommand{\TPC}          {\rm{TPC}\xspace}
\newcommand{\TRD}          {\rm{TRD}\xspace}
\newcommand{\VZERO}        {\rm{V0}\xspace}
\newcommand{\VZEROA}       {\rm{V0A}\xspace}
\newcommand{\VZEROC}       {\rm{V0C}\xspace}
\newcommand{\Vdecay} 	   {\ensuremath{V^{0}}\xspace}

\newcommand{\ee}           {\ensuremath{e^{+}e^{-}}} 
\newcommand{\pip}          {\ensuremath{\pi^{+}}\xspace}
\newcommand{\pim}          {\ensuremath{\pi^{-}}\xspace}
\newcommand{\kap}          {\ensuremath{\rm{K}^{+}}\xspace}
\newcommand{\kam}          {\ensuremath{\rm{K}^{-}}\xspace}
\newcommand{\pbar}         {\ensuremath{\rm\overline{p}}\xspace}
\newcommand{\kzero}        {\ensuremath{{\rm K}^{0}_{\rm{S}}}\xspace}
\newcommand{\lmb}          {\ensuremath{\Lambda}\xspace}
\newcommand{\almb}         {\ensuremath{\overline{\Lambda}}\xspace}
\newcommand{\Om}           {\ensuremath{\Omega^-}\xspace}
\newcommand{\Mo}           {\ensuremath{\overline{\Omega}^+}\xspace}
\newcommand{\X}            {\ensuremath{\Xi^-}\xspace}
\newcommand{\Ix}           {\ensuremath{\overline{\Xi}^+}\xspace}
\newcommand{\Xis}          {\ensuremath{\Xi^{\pm}}\xspace}
\newcommand{\Oms}          {\ensuremath{\Omega^{\pm}}\xspace}
\newcommand{\degree}       {\ensuremath{^{\rm o}}\xspace}

\begin{titlepage}

\title{Sensor operating point calibration and monitoring of the ALICE Inner Tracking System during LHC Run 3}
\ShortTitle{Calibration and monitoring of the ALICE ITS2 in  Run 3}   

\Collaboration{ALICE ITS Collaboration\thanks{See Appendix~\ref{app:collab} for the list of collaboration members}}
\ShortAuthor{ALICE ITS Collaboration} 

\begin{abstract}
The new Inner Tracking System (ITS2) of the ALICE experiment began operation in 2021 with the start of LHC Run 3. Compared to its predecessor, ITS2 offers substantial improvements in pointing resolution, tracking efficiency at low transverse momenta, and readout-rate capabilities.
The detector employs silicon Monolithic Active Pixel Sensors (MAPS) featuring a pixel size of 26.88$\times$29.24~$\upmu$m$^2$ and an intrinsic spatial resolution of approximately 5~$\upmu$m. With a remarkably low material budget of 0.36\% of radiation length ($X_{0}$) per layer in the three innermost layers and a total sensitive area of about 10 m$^2$, the ITS2 constitutes the largest-scale application of MAPS technology in a high-energy physics experiment and the first of its kind operated at the LHC.
For stable data taking, it is crucial to calibrate different parameters of the detector, such as in-pixel charge thresholds and the masking of noisy pixels. The calibration of 24120 monolithic sensors, comprising a total of 12.6$\times$10$^{9}$ pixels, represents a major operational challenge. This paper presents the methods developed for the calibration of the ITS2 and outlines the strategies for monitoring and dynamically adjusting the detector’s key performance parameters over time.

\end{abstract}

\end{titlepage}

\setcounter{page}{2} 



\section{Introduction}\label{sec:intro}
The Inner Tracking System (ITS) of the ALICE experiment has been replaced in 2021 by a completely new detector, called ITS2, to fully exploit the luminosity of the CERN Large Hadron Collider (LHC) during Runs 3 and 4.
The installation of the new detector was also made possible by the installation of a new beam pipe, featuring a central beryllium section with an outer radius reduced from 28~mm to 18~mm ~\cite{ALICE:2023udb}. 
The new tracker is composed of seven cylindrical and concentric layers equipped with 24120 silicon Monolithic Active Pixel Sensors (MAPS) called ALPIDE~\cite{ALICE:2023udb, AglieriRinella:2017lym}, covering a total active area of about 10~m$^2$. 
It has a low material budget of 0.36\% $X_{0}$/layer in the innermost layers, and it significantly enhances the charged-particle pointing resolution and standalone tracking efficiency down to very low transverse momenta compared to the previous tracker~\cite{ALICE:2013nwm,ALICE:2023udb}. 
The sensor (hereafter called \textit{chip}), produced by TowerJazz~\cite{tower} with the 0.18~$\upmu$m CMOS imaging process technology, features a pixel size of 26.88$\times$29.24 $\upmu$m$^{2}$ with a total of 512$\times$1024 pixels distributed over 15$\times$30~mm$^{2}$. This translates into a total number of 12.6 billion pixels in the full detector. The sensor features a binary readout, with the signal discriminated in-pixel.

The detector is divided into an Inner Barrel (IB), with three layers, and an Outer Barrel (OB) with four layers (two Middle Layers, ML, and two Outer Layers, OL), and it is segmented into 192 Staves with a maximum length of about 148 cm (OL staves). 
Each IB stave is formed by 9 chips with independent high-speed data links with 1.2 Gb/s bandwidth each. The power to the chips is supplied directly through a Flexible Printed Circuit (FPC) wire-bonded to the sensors. For the OB staves, a single high-speed data link at 400 Mb/s allows the readout of groups of 7 chips each. The power supply is provided through external power and bias buses soldered to the FPCs. See Fig.~23 and 24 of Ref.~\cite{ALICE:2023udb} for more details. 
The Hybrid Integrated Circuit (HIC), also called the \textit{module}, comprises the group of chips interconnected by data links: one group for the IB and two groups for the OB. 
There are no other active components than the ALPIDE chips mounted on the detector staves. The detector barrels are cooled using a water cooling system that operates at sub-atmospheric pressure with a constant water flow and temperature.

The readout of the detector can be either triggered or continuous. In the continuous mode, equally spaced triggers are continuously sent to the sensors, and the data are recorded for the full duration of each trigger signal (readout window). The trigger frequency, also called \textit{framing rate} in the following, generally varies between 11 and 202 kHz. In the triggered mode, pixel hits on the sensors are latched into the sensor memory and read out only if a physics trigger is sent from the ALICE Central Trigger Processor (CTP)~\cite{Krivda:2016myl} within a few microseconds after the event that generated them. This limits the readout to specific events. 

The power and reverse bias voltages needed to operate the chips are provided by external, custom Power Boards (PBs). The PBs are the last stage of active regulation and are connected via approximately 7\,m long cables to the detector. The resulting voltage drops have to be compensated for at the level of the PB. On IB staves, the return paths for analogue and digital domains are separated. On OB staves, the analogue and digital power domains of all modules instead share the same return path (digital and analogue grounds are connected at the level of the power bus on the stave), leading to a more complex interplay of the modules.
The voltage drop is compensated using an iterative procedure that begins with an initial output voltage of 1.8 V at the PB, taking into account the known resistance of the power cables. In each iteration, the digital and analogue currents are read from the PB, the corresponding voltage drop across the cables is calculated, and this drop is added to the 1.8 V reference. This results in an increased PB output voltage for the next iteration. It was found that four iterations are sufficient to regulate the voltage on the chips with a precision of $\pm$35 mV around the target value of 1.8 V. The voltage drop on the HICs is negligible compared to that introduced by the long power cables. Voltage stability at the chip level is ensured by several decoupling capacitors mounted close to the chips. With the ITS2 powering scheme, voltage drop compensation is applied individually to each stave in the IB and to each HIC in the OB. Typical supply voltages at the PB range from 1.9 to 2.1 V, or 2.1 to 2.3 V for analogue and digital domains, respectively, depending on the stave. For standard physics data taking, the reverse bias voltage is kept at 0 V, but it can be reduced to a negative value if measurable radiation damage effects are observed\footnote{No measurable radiation damage effects were observed up to the time of final manuscript preparation (December 2025).}. 
The large number of channels and the complex power distribution scheme make the monitoring and calibration of the detector a crucial and challenging task to be performed before recording physics data to ensure stable data taking.

The detector calibration consists of tuning the in-pixel thresholds to a target value at which the sensors are fully efficient while maintaining a low fake-hit rate, and of masking noisy pixels. The thresholds of the pixels are tunable chip by chip, and they are typically set in the range between 100 and 150~$e^{-}$. With thresholds in this range and for Minimum-Ionizing Particles (MIP), test beam results show that the sensors achieve an intrinsic spatial resolution of 5 $\upmu$m, a mean cluster size ranging between 2 and 3 pixels, and a detection efficiency above 99\%, with a fake-hit rate below 10$^{-6}$ hits/event/pixel when only a negligible number of pixels are masked~\cite{ALICE:2023udb}. This entirely fulfills the design requirements of ITS2~\cite{ALICE:2013nwm}. The design requirement for the fake-hit rate is crucial to limit combinatorics during the track reconstruction~\cite{Buncic:2015ari}.
In addition, the detector working parameters need to be monitored daily or yearly (depending on the parameter) to check their stability over time. A series of tests (also called \textit{scans}) were developed for this purpose. 
In general, calibration and monitoring procedures are performed using the ALICE computing farm, which also serves for physics data taking and is shared across the full experiment. 
Another calibration procedure performed on the ALPIDE chip involves its internal temperature sensor output. Details of this procedure are provided in Appendix~\ref{appendix:temp}.

\section{ALPIDE chip functionalities for sensor calibration}\label{sec:alpide}
\begin{figure}
    \centering
    \includegraphics[width=0.65\textwidth]{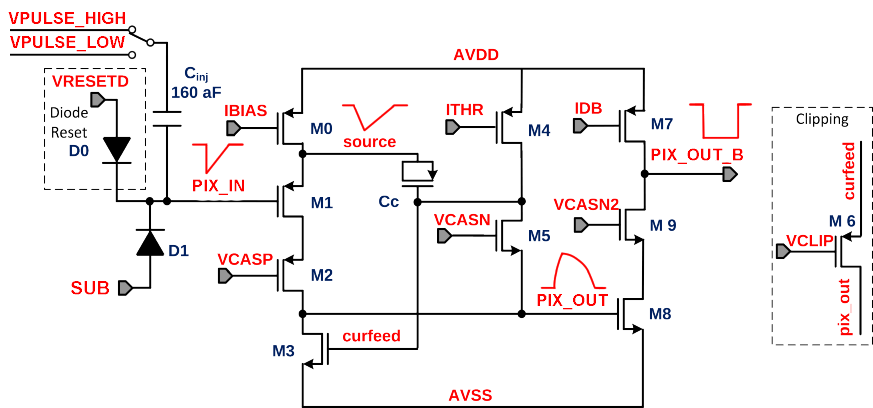}
    \caption{Scheme of the ALPIDE in-pixel analogue front-end. Four signal sketches are included to illustrate how the signal is shaped at the different stages.}
    \label{fig:alpideanalogue}
\end{figure}

This section provides a short description of the ALPIDE functionalities connected to the calibration results discussed in this article. For a more complete description, see Sec.~2.3 in Ref.~\cite{ALICE:2023udb}, Ref.~\cite{Kim:2016ktw}, and Ref.~\cite{AglieriRinella:2017lym}. \\ \\ 
The pixel matrix of the ALPIDE chip is arranged into double columns since pairs of adjacent columns share the same \textit{priority-encoder} based readout circuit \cite{Yang:2015vpa}.
Each pixel contains an analogue front-end circuit amplifying, shaping, and discriminating the signal collected by the sensing node, as well as digital circuitry. On the long edge of the chip, a periphery circuit region of 1.2 $\times$ 30 mm$^{2}$ includes all the readout and control functionalities.
\\ \\ 
Figure~\ref{fig:alpideanalogue} shows the schematic of the analogue front-end circuit. The analogue voltage and its ground are indicated as AVDD and AVSS, respectively. 
The diode \texttt{D1} is the sensor p-n junction. The sensing node is continuously reset by the diode \texttt{D0} and \texttt{VRESETD} establishes its reset voltage. The reset current depends exponentially on the forward bias of the diode \texttt{D0}.
The potential at \texttt{PIX\_IN} depends on \texttt{VRESETD} and on the voltage drop across the diode \texttt{D0} due to the leakage current in the pixel.
A particle hit will lower the potential at the pixel input \texttt{PIX\_IN} by a few tens of millivolts\footnote{The input capacitance is of the order of a few fF, assuming a collected charge of a few hundreds of electrons.}.
In this case, the \textit{source} node is forced by the source follower (formed by \texttt{M1} and the current source \texttt{M0}) to follow the voltage excursion, transferring the associated charge onto the analog output node \texttt{PIX\_OUT}. Simultaneously, the coupling between the source node and the \textit{curfeed} node reduces the current in \texttt{M3}. These two effects combine to raise the potential at \texttt{PIX\_OUT} by several hundred millivolts, forcing \texttt{M8} into conduction. If the charge deposited by the particle hit is large enough to overcome the current set by \texttt{IDB} on \texttt{M7}, \texttt{M8} will pull the \texttt{PIX\_OUT\_B} node down to zero.
The charge threshold of the pixel can be adjusted through \texttt{ITHR} and \texttt{VCASN}. The increase of \texttt{VCASN} causes an exponential decrease of the charge threshold (coarse threshold tuning) while the increase of \texttt{ITHR} generates a linear increase of the threshold, allowing for a finer adjustment. 
The cascode transistor \texttt{M9} reduces the equivalent Miller capacitance
on \texttt{PIX\_OUT}.
The voltage bias \texttt{VCLIP} controls the gate of the clipping transistor \texttt{M6}, limiting the maximum excursion of the analogue signal.
The lower \texttt{VCLIP} is, the sooner the signal clipping will set in.
With the default setting of maximum clipping (\texttt{VCLIP} = 0), the output of the front-end has a peaking time of the order of 2 $\upmu$s, while the discriminated pulse has a typical duration of 6--8 $\upmu$s. This feature is particularly important in continuous-mode readout, as limiting the signal duration to 6--8 $\upmu$s prevents the same pixel from being read out in multiple consecutive triggers of similar duration.
\begin{figure}[h!]
    \centering
    \includegraphics[width=0.85\textwidth]{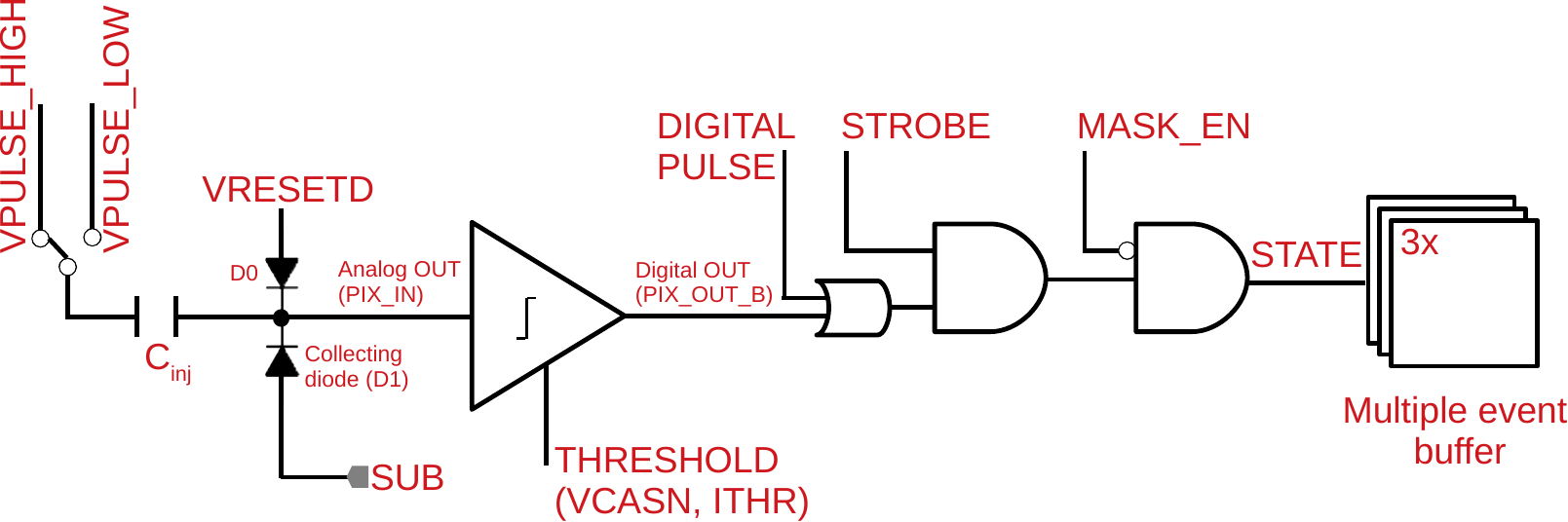}
    \caption{Simplified schematic view of the ALPIDE pixel cell.} 
    \label{fig:alpidedigital}
\end{figure}

For calibration purposes, an injection capacitor $\mathtt{C_{\rm inj}}$ allows the injection of a test charge into the input of the front-end (analogue pulse). The voltage step is controlled via on-chip biasing Digital-to-Analogue Converters (DACs) in the chip periphery region. The amplitude of the applied voltage pulse is defined by the difference between \texttt{VPULSE\_HIGH} and \texttt{VPULSE\_LOW}, in DAC units, as shown in Fig.~\ref{fig:alpideanalogue}. The two edges of the pulse provoke the injection of two charge pulses of opposite polarities. The leading falling edge of the analogue pulse corresponds to the discharge of the collection diode, in a manner equivalent to the passage of a charged particle.
One DAC unit corresponds to 10 $e^-$ based on the design value and parameters of the injection capacitor and DACs. Throughout this article, this conversion factor is used to convert DAC units to electrons. \texttt{V\texttt{PULSE}\_HIGH} and \texttt{V\texttt{PULSE}\_LOW} can range between 0 and 170 DAC units, resulting in a maximum injectable charge of 1700 $e^-$. The analogue circuitry is summarised in the left part of Fig.~\ref{fig:alpidedigital} up to the discriminator.
\\ \\ 
The digital pixel circuitry features a multi-event buffer capable of storing up to three hits, along with a pixel mask register. A digital-only pulsing (\texttt{DIGITAL PULSE}) is available to activate the in-pixel hit register, bypassing the discriminator. The latching of the discriminated hit in one of the three hit storage registers is controlled by a global \texttt{STROBE} signal. The global \texttt{STROBE} signal addresses the individual hit registers in a round-robin fashion, using individually routed lines for each of the three hit storage registers. During normal operations, all three hit storage registers are used.
A pixel hit is set when there is a coincidence between the discriminated signal and the \texttt{STROBE} as shown in Fig.~\ref{fig:alpidedigital}. For calibration and monitoring scans, the strobe and pulse signals are generated through the internal ALPIDE sequencer. This is activated through an external trigger source, as will be outlined in Sec.~\ref{sec:tech}. The logic also provides the possibility of masking the pixel output. When the control bit \texttt{MASK\_EN} is set high, the \texttt{STATE} output is forced to 0, effectively masking the pixel output to the priority encoder. The low value provides normal functionality. In addition, a set of registers in the chip periphery can be used to disable double columns (this disables the priority encoder, in case of issues with the pixel masking). The priority encoder of a double column automatically switches off when a stuck pixel is detected, i.e., a pixel that sends its address to the chip periphery twice in succession. 
The duration of the \texttt{STROBE}, digital, and analogue pulses, as well as their relative timing, are programmable. The typical duration of the digital and analogue pulses is on the order of a few tens of $\upmu$s, while the \texttt{STROBE} signal usually lasts between 25 ns and approximately 10~$\upmu$s. The latter can be delayed with respect to the pulse (simply called \textit{strobe delay}), with typical delays ranging from 25 ns to 10~$\upmu$s. When maximum signal clipping is applied, the duration of \texttt{PIX\_OUT\_B} is constrained to about 6--8 $\upmu$s. In addition, a significantly lower \texttt{ITHR}, compared to its nominal value of 50 DAC units, results in a longer pulse duration, which changes the time spent by the signal above the charge threshold. \\ \\
The periphery of the chip contains 14 8-bit analogue DACs for the biasing of the pixel front-end circuitry (namely, \texttt{ITHR}, \texttt{VCASN}, \texttt{IDB}, etc.). The biases are applied chip-wise without the possibility of tuning individual pixels or sub-groups of pixels. The pixel-by-pixel variations solely depend on CMOS manufacturing tolerances and on the presence of pads on the chip (allowing for the chip interconnection with the external FPC) that slightly interfere with the pulsing process. 
The chip periphery also contains monitoring circuitry, namely a 10-bit Analogue-to-Digital Converter (ADC) and a temperature sensor, as well as a band-gap voltage reference. This is the specific voltage reference for the ADC (DACs are referred to AVDD-AVSS).
The ADC can be used to monitor internal analogue signals like voltage and current outputs of the internal DACs, as well as analogue and digital supply voltages, and to read out the temperature sensor. 
The ADC is composed of four main blocks: an input scaling stage to adapt the analogue output of the DACs to the ADC input dynamic range, an ADC-internal DAC that generates a voltage ramp, a comparator that compares the input analogue signal with the voltage ramp, and a digital control block. 
The dynamic range of the ADC-internal DAC is limited, which results in a limitation of the ADC input dynamic range to about 1.72 V. For this reason, the AVDD measurement is performed through the voltage DAC \texttt{VTEMP}, which acts as a voltage divider between AVDD and its ground. The voltage DAC \texttt{VTEMP} is not used to bias the pixel matrix but specifically for the indirect measurement of AVDD which is a prerequisite for the ALPIDE temperature sensor calibration, as illustrated in detail in Appendix \ref{appendix:temp}.

\section{Calibration and monitoring scans}\label{sec:scans}
Calibration and monitoring scans are employed to tune and monitor the working point of the detector (charge thresholds and noisy-pixel masks) and to monitor its working conditions (\texttt{VRESETD} scan and pulse-shape scan). The working point is chosen to optimize detection efficiency while keeping the fake-hit rate at an acceptably low level.
In general, some scans are performed through pixel activation using pulsing, while other scans are conducted without any stimuli.
The first category contains the analogue, threshold, \texttt{VCASN}, \texttt{ITHR}, \texttt{VRESETD} and the pulse-shape scans.
The other scans without stimuli are the digital scan, which relies on the digital activation of the pixel latches (multi-event buffer), and the noise scan, which exploits a data taking run in absence of beam collisions. 
These scans are explained in detail in this section. All scans are performed with a reverse substrate bias voltage of 0 V, corresponding to the voltage used in standard physics data taking. Table~\ref{tab:scans} summarizes the list of scans along with their objective, their execution frequency, and a brief description. In this table, the threshold scan is considered among the monitoring scans; however, it represents a fundamental scan to verify the tuning of the thresholds.
%
\begin{table*}[th!]
\caption{Summary of all calibration and monitoring scans of ITS2.}
  \label{tab:scans}
  \noindent
  \centering
  \begin{tabular}{lccl}
  	\hline
  	Scan & Objective & Frequency & Description \\
    \hline
      Digital/Analogue & Calibration & $\sim$1/year & Tagging of problematic pixels and pixel columns \\
      \texttt{VCASN} & Calibration & $\sim$1/year & Coarse tuning of pixel thresholds \\
      \texttt{ITHR} & Calibration & $\sim$1/year & Fine tuning of pixel thresholds \\
      Short Threshold & Monitoring & $\sim$1/day & Measurement and monitoring of pixel thresholds \\
      Full threshold & Monitoring & 1/year & Measurement of pixel thresholds used as reference \\
      \texttt{VRESETD} & Monitoring & 1/year & Monitoring of \texttt{VRESETD} working point \\
      Pulse shape & Monitoring & $\sim$1/year & Study of the pixel analogue signal characteristics\\
      Noise & Calibration & $\sim$1/year & Detection of noisy pixels\\
    \hline
  \end{tabular}
\end{table*}
\subsection{Digital and analogue scans}
The \texttt{digital} and \texttt{analogue scans} are used to check the response of the in-pixel circuitry and the \textit{priority encoder}.
The \texttt{analogue scan} probes both the analogue front-end and digital in-pixel circuitry, while the \texttt{digital scan} is limited to the digital in-pixel circuitry.
For the \texttt{digital scan}, each pixel is activated 50 times. For the \texttt{analogue scan}, a fixed test charge above the mean in-pixel charge threshold is injected 50 times in each pixel. In both cases, injections are performed row by row (in all ALPIDE rows) and the chip matrix is read out to check the pixel response. Stuck pixels are not masked in this procedure. 
These scans allow the identification of \textit{dead} pixels (pixels without hits), \textit{inefficient} pixels (pixels with less than 50 hits), and $noisy$ pixels (pixels with more than 50 hits). If a double-column (1024 pixels) contains more than 50 noisy pixels, the issue is typically assigned to a problem in the \textit{priority encoder} circuitry that reads out the double column. This often disrupts either the propagation of the pixel address information to the chip periphery or the reset signal from the periphery to the pixel. Since such double columns are found to generate excessive amounts of data and occasionally stall the readout, they are excluded from the readout and data taking (disabled through the double column registers, see Sec.~\ref{sec:alpide}). The frequency of these cases is discussed in Sec.~\ref{sec:noisydcols}.

\subsection{VCASN, ITHR, and Threshold scans}\label{scurveScans}
\begin{figure}[h!]
    \centering
    \includegraphics[width=0.6\textwidth]{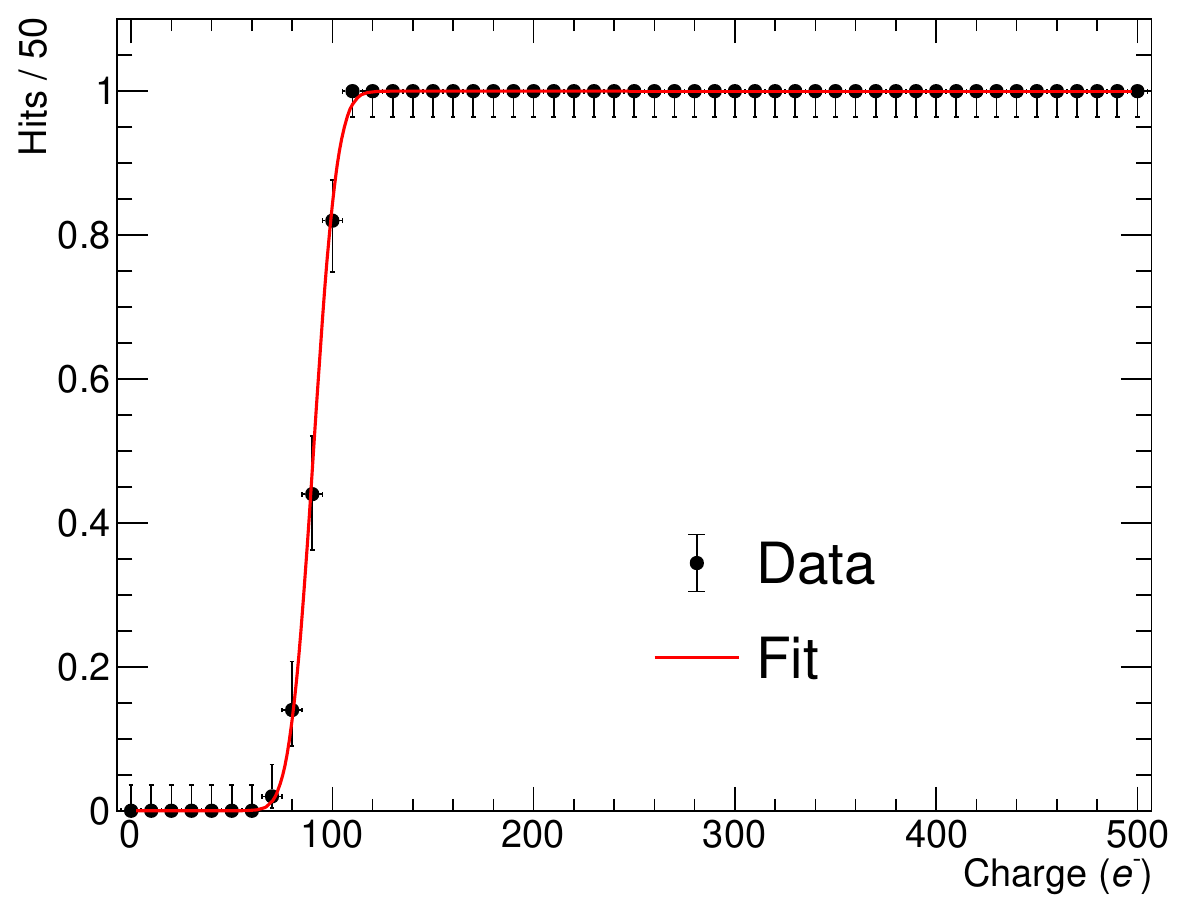}
    \caption[sc]{Number of hits normalized to the number of injections per charge point as a function of the injected charge (in electrons) for a pixel of ITS2 in a threshold scan. The errors are evaluated as Binomial errors based on the Clopper--Pearson~\cite{Clopper:1934sws} method with a confidence level of 68.27\%, as provided by the ROOT \texttt{TEfficiency} class~\cite{tefficiency}. The red line represents a fit to the data performed with an error function. See text for more details.}
    \label{fig:scurve}
\end{figure}

This group of scans relies on repeated charge injections as a function of a scan parameter and on the measurement of the number of hits registered for each pixel per setting. In all scans, injections are performed row by row. The external trigger frequency is set to 1 kHz, with an ALPIDE pulse and strobe duration of 12.5 $\upmu$s and 2 $\upmu$s, respectively.  
In the \texttt{threshold scan}, the analogue pulsing is repeated 50 times for each test charge ranging from 0 to 500~$e^-$ in steps of 10~$e^-$. In the \texttt{VCASN scan}, the \texttt{VCASN} biasing parameter of the ALPIDE front-end is varied from 30 to 70 DAC units
in steps of 1 DAC unit while keeping the other biasing parameters to default design values and injecting a constant charge equal to the target threshold 50 times per setting. The same procedure is followed in the \texttt{ITHR scan} with the difference that the \texttt{ITHR} biasing parameter is varied from 30 to 100 DAC.
The tuning of the thresholds is of crucial importance to keep the ALPIDE in a threshold regime between 100 and 150~$e^-$ in which the ALPIDE has a detection efficiency above 99\% and a fake-hit rate lower than 10$^{-6}$ hits/event/pixel (see Fig. 12 of Ref.~\cite{ALICE:2023udb}). 
\\ \\
The number of hits as a function of the scan parameter generates an \textit{S-curve} for every pixel, as shown in Fig.~\ref{fig:scurve} for a \texttt{threshold scan} of a pixel in the detector. In principle, a fit with an error function\footnote{Fit equation:  $(N_{\rm inj} / 2)\times \left(1 + \mathrm{erf} \left(\frac{x-\mu}{\sqrt{2}\sigma}\right)\right)$ where $\mathrm{erf} \left(y\right) = 2/\sqrt{\pi} \int_{0}^{y} e^{-x^{2}} \,dx$ is the error function, $\mu$ is the 50\% point which represents the threshold, $\sigma$ is the slope of the S-curve which represents the temporal noise, and $N_{\rm inj}$ is the number of charge injections per point, which corresponds to 50. Only for the \texttt{ITHR} scan, the S-curve is mirrored compared to the other scans, hence a minus sign between the 1 and the $\mathrm{erf}()$ is needed.} can be performed to the data to extract the 50\% point, which is taken as an estimate of the threshold, and the slope around this point, which is considered to be the temporal noise. However, a fit based on \texttt{ROOT} \texttt{TMinuit}~\cite{James:2296388} takes about 1~ms per pixel (see Sec.~\ref{sec:tech} for details on the computing farm architecture). To drastically reduce the processing time to the level of 1--2 ms per row (1024 pixels), the threshold and noise are extracted by calculating the numerical derivative of the data. 
The numerical derivative is obtained from the difference between adjacent bin contents in Fig.~\ref{fig:scurve}: for each charge value $c$, the difference between the number of hits at $c+1$ and at $c$ is assigned.
The resulting numbers will be distributed according to a Gaussian distribution: its mean and standard deviation are directly the threshold and the temporal noise, respectively. This method takes about 1--2 $\upmu$s per pixel because it executes the simple calculation of the mean and standard deviation of a set of numbers. For \texttt{VCASN} and \texttt{ITHR} scans, the $x$ axis of Fig.~\ref{fig:scurve} will contain the values of the two registers, respectively, and the 50\% point will be the DAC setting at which the pixel responded to the injection of the charge equal to the target threshold. 
\\ \\
For all scans, the parameters of interest are the mean and the standard deviation of the distribution of the measured quantities over all pixels of a chip (e.g. mean chip threshold, mean \texttt{VCASN}, mean \texttt{ITHR}, etc.).
As ALPIDE uses a global biasing scheme, without tuning of individual or groups of pixels, and since it was found that a small fraction of the pixels is representative for the whole pixel matrix, it is sufficient to scan only a small percentage of pixels to assess the full chip. This correspondence has been established during the characterization of single ALPIDE chips in the laboratory and has been verified during commissioning, as well as after the first year of data taking.
As a consequence, the \texttt{threshold scan} is typically run on  2.1\% of the pixels. This version of the scan is called \texttt{short threshold scan} in the following. The \texttt{VCASN} and \texttt{ITHR scans} run on an even lower fraction of 0.7\% of the pixels. These pixels are arranged in rows that are uniformly distributed across the full matrix, while avoiding placement beneath areas where the chip is interconnected to the external FPC, as explained in Sec.~\ref{sec:alpide}. However, a \texttt{full threshold scan} (run on all rows) is entirely feasible, and it is performed about once per year as a reference.\\ \\
In order to achieve the target threshold, a \texttt{VCASN scan} and an \texttt{ITHR scan} are executed in sequence, starting from the \texttt{VCASN} scan and keeping the \texttt{ITHR} to 50 DAC counts for all chips. Then, during the \texttt{ITHR scan}, the previously determined \texttt{VCASN} values are used. As briefly discussed in Sec.~\ref{sec:alpide}, the finer adjustment performed with \texttt{ITHR} ensures that the pulse duration is not significantly modified by the tuning procedure.
Each new tuning is validated with a \texttt{short threshold scan}. Moreover, the \texttt{short threshold scan} is also used as a monitoring tool to measure the pixel thresholds after every LHC fill, which allows to assess the time stability of the threshold setting. 

\subsection{VRESETD scan}\label{sec:vresetd-scan}
The \texttt{VRESETD scan} is used to check the operational reset voltage range for a desired threshold. As further discussed in Sec.~\ref{subsec:resultsVRESETD}, the \texttt{VRESETD} working point gets modified with accumulated radiation dose, hence it is important to perform a \texttt{VRESETD} scan at least once a year to monitor its working point\footnote{With particular attention to the end-of-the-year Pb--Pb collisions where most of the radiation damage occurs.}.
Two versions of this scan exist: 
\begin{itemize}
    \item \texttt{VRESETD uni-dimensional scan}: the \texttt{VRESETD} biasing parameter is varied from 100 to 240~DAC units in steps of 5 DAC units. For each \texttt{VRESETD} value, a fixed charge corresponding to 300~$e^-$, well above the in-pixel threshold, is injected. The same row in each chip is used for this scan. This scan allows for a coarse determination of the operating margins of \texttt{VRESETD}.
    \item \texttt{VRESETD two-dimensional scan}: the \texttt{VRESETD} biasing parameter is varied from 100 to 240 DAC units in steps of 5 DAC units. The difference with respect to the uni-dimensional scan is that for every \texttt{VRESETD} value a \texttt{threshold scan} is performed. In this scan, the dependence of the threshold on the reset voltage is mapped out at the cost of longer execution time.
\end{itemize}
A single row is selected for both scans primarily to reduce the execution time. As explained in Sec.~\ref{scurveScans}, the chosen row also avoids crossing ALPIDE pads, which could slightly interfere with the measurements. In both versions, the external trigger frequency is set to 1 kHz, with an ALPIDE pulse and strobe duration of 12.5 $\upmu$s and 2 $\upmu$s, respectively.

\subsection{Pulse-shape scan}\label{sec:pulseshape}
The \texttt{pulse-shape scan} is used to probe the analogue front-end pulse shape of a representative set of pixels (1 row) and to measure how the signal is modified by the clipping DAC \texttt{VCLIP} and by \texttt{ITHR}. The usage of a single row ensures a reasonable duration of the scan. As explained in Sec.~\ref{scurveScans}, the chosen row also avoids crossing ALPIDE pads, which could slightly interfere with the measurements. 
By having a small strobe duration compared to the pulse duration, and by shifting in time the propagation of the strobe, it is possible to probe the ALPIDE analogue signal shape as a function of the strobe delay. The external trigger frequency is set to 1 kHz. 
This scan exists in two versions:
\begin{itemize}
    \item \texttt{Pulse shape uni-dimensional scan}: the delay between the analogue pulse and the internal strobe, called \textit{strobe delay}, is varied from 0 to 10 $\upmu$s in steps of 25~ns. The analogue pulse length is set to 12.5~$\upmu$s to have the rising edge outside the strobe window. The width of the strobe signal is set to 25~ns. For every value of the strobe delay, a fixed charge corresponding to 300 $e^-$ is injected.
    \item \texttt{Pulse shape two-dimensional scan}: the strobe delay is varied from 0 to 50 $\upmu$s in steps of 250~ns. The analogue pulse length is set to 62.5~$\upmu$s and the width of the strobe signal is set to 250 ns. For every value of the strobe delay, a \texttt{threshold scan} is performed, but unlike the standard threshold scan described in Sec. \ref{scurveScans}, the charge injected ranges from 0 to 1700 $e^-$. In this way, it is possible to probe the ALPIDE signal shape as a function of the injected charge (see also Fig.~\ref{fig:alpidedigital}).
\end{itemize}

With the ALPIDE settings mentioned above, the uni-dimensional scan provides an accurate estimate of the pulse parameters compared to the two-dimensional scan, thanks to the finer strobe delay steps. The \texttt{two-dimensional pulse shape scan} enables mapping of the same parameters as a function of the injected charge. Results from these scans are shown in Sec.~\ref{sec:results}.

\subsection{Noise scan}\label{sec:noise-calib}
The noise scan is performed without analogue or digital injections, and consists of a run that can be recorded at different framing rates between 11 and 202 kHz in the absence of beam collisions. The detector is read out in the continuous mode without any charge injection. In every readout frame, the fired pixels from every chip are recorded as in a standard physics run. At the end of the scan, the number of hits recorded on the fired pixels is stored. This allows the calculation of the per-pixel fake-hit rate, which is used to tag noisy pixels. To ensure a precise estimation of the per-pixel fake-hit rate, the minimum required number of events is defined as the smallest integer greater than or equal to $(1+1/t)/\delta^{2}$, where $t$ denotes the pixel occupancy threshold and $\delta=0.2$ represents a maximum relative error of 20\% on $t$\footnote{Provided that the pixel fired in $n$ out of $N$ readout frames, the occupancy is given by $t = n/N$. The relative error on $t$ follows from Poissonian statistics $(\sigma_{\mathrm{t}} / t)^{2} = 1/n + 1/N = 1/N(1+1/t)$. Given that the relative error on the occupancy is required to be less than some chosen $\delta$, then $N > (1+1/t)/\delta^{2}$.}. In the OB layers, the threshold $t$ is set to 10$^{-6}$ hits per readout frame (event). This means that every pixel providing, on average, more than 10$^{-6}$ hits per frame is considered noisy and hence masked during data taking. For the IB chips, a looser cut $t$ of 10$^{-2}$ hits per frame is applied: given the smaller number of channels with respect to the OB, each chip can exploit more bandwidth. This also leads to a maximization of the detection efficiency. Given the defined values of the cut $t$, 2525 events are enough for the IB, while for the OB, 25'000'025 events are needed. The recording and processing of these events takes approximately 15--20 minutes with a framing rate of 67 kHz, considering bottlenecks related to the data processing (see Sec.~\ref{sec:tech}). The variation over time of the fake-hit rate is monitored on a monthly basis with cosmic-ray runs recorded during periods with no beam collisions (technical stops, machine interventions, etc.). They are purely used to monitor the fake-hit rate and not to mask new noisy pixels. The pixel masks are renewed approximately every year, as reported in Tab.~\ref{tab:scans}). 

\section{Computing and software infrastructure chain for calibration scans}\label{sec:tech}
\begin{figure}
    \centering
    \includegraphics[width=0.9\linewidth]{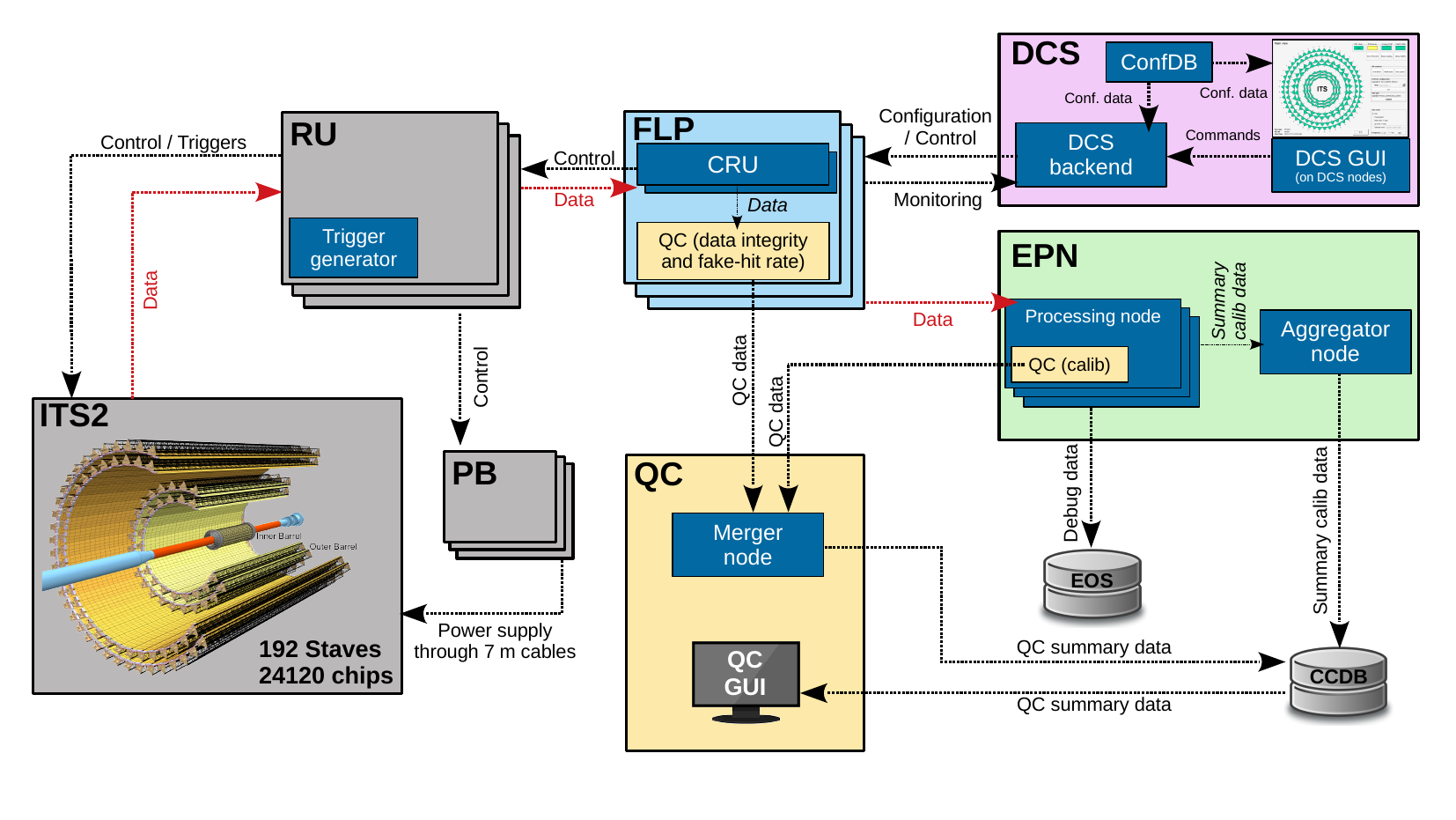}
    \caption{Sketch of the ITS2 hardware and software chain needed to run and analyse calibration scans.}
    \label{fig:fullscheme}
\end{figure}

This section outlines in detail the full software and hardware chain that are used to run, analyse and store the results of the ITS2 calibration scans described in Sec.~\ref{sec:scans}. 
The full chain is sketched in Fig.~\ref{fig:fullscheme}. 
Starting from the left side of Fig.~\ref{fig:fullscheme}, the ITS2 Staves are read out by 192 identical Readout Units (RU) which provide control and trigger signals, and read out the high-speed data lines from the ALPIDE chips. The power supply is provided through custom PBs, controlled through the RUs, as explained in Sec.~\ref{sec:intro}. In calibration scans, the triggers are generated by the RUs themselves by means of an internal programmable \textit{sequencer}. This allows the detector to be completely independent of the external trigger provided by the ALICE Central Trigger Processor (CTP) \cite{Krivda:2016myl}, and also allows scans to be performed independently on different staves. The CTP trigger is used only in noise scans.
The 192 RUs are connected through GigaBit Transceiver (GBT) links to 22 custom FPGA-based Common Readout Units (CRU) which are mounted into 13 First Level Processors (FLPs) located about 100~m from the detector, in an area with no radiation exposure. More details can be found in Refs. \cite{ALICE:2023udb} and \cite{Buncic:2015ari}. 
The FLPs communicate with a Detector Control System (DCS) node, referred to as the DCS \textit{backend}, through a shared network. They exchange DCS data with the RUs (and PBs) via their associated CRUs.\\ \\
The DCS software of the ITS2, represented by a pink box in Fig.~\ref{fig:fullscheme}, consists of three main components: a WinCC Graphical User Interface (GUI), a low-level C++ backend software (\textit{ITSComm}) and a Configuration Database (\textit{ConfDB}). The actual configuration of the detector and thus also the execution of the scans is performed by \textit{ITSComm}, based on the settings stored in the configuration database. 
The configuration database contains the current configuration settings of the chips and the RUs, as well as the calibration parameters for PBs (for the voltage drop correction). They are accessed by ITSComm to configure the detector and its services, as well as by WinCC panels for consultation. 
\\ \\
The 13 FLPs of ITS2 are connected to a farm of 350 Event Processing Nodes (EPN) shared among all the ALICE detectors (see Sec. 5.3.1 in Ref.~\cite{ALICE:2023udb} for more details on the farm). They are sketched in Fig.~\ref{fig:fullscheme} with a green box.
The ALICE Offline and Online System (O$^{2}$) \cite{Buncic:2015ari, ALICE:2023udb} software allows processing and analysis of the ITS2 calibration data on the farm. Following the red arrows in Fig.~\ref{fig:fullscheme}, the raw data arriving on the EPNs are organized into Time Frames (TFs)  with a duration of 2.85 ms each. The processing, analysis and amount of data on the EPNs depend on the type of calibration scan. \\ \\
Regarding the data rate, the CTP trigger frequency during noise scans typically ranges from 11 to 202~kHz. For a typical frequency of 67 kHz, the readout rate of the full detector is about 8 GB/s (sum of all FLP's readout rates). All other calibration scans, which make use of the RU internal sequencer, are typically recorded at a trigger frequency of 1 kHz. Supposing that a full pixel row (1024 pixels) fires all the time during a scan, this would result in a total data rate at the FLPs of 37 GB/s\footnote{The hit of two neighbouring pixels in a double column is transmitted with a single word of 3 bytes. If a full row fires, 512 words of 3 bytes each are sent out by the ALPIDE chip.}. Considering the overhead introduced by changing the chip settings (e.g., change of injected charge, change of row, etc.) and by the FLPs in sending the commands to the sensors, the data rate never exceeds 9 GB/s. As a reference, this overhead extends the duration of a short threshold scan executed at 1~kHz from 28 to approximately 120 seconds. Overall, a short threshold scan corresponds to about 1 TB of raw data while a full threshold scan (the most demanding among all calibration scans) is equivalent to about 48~TB\footnote{This supposes that pixels always fire in a threshold scan. This number reduces to about 39 TB in the ideal case that all pixel thresholds are equal to 100~$e^-$.}.\\ \\
The decoding of ALPIDE raw data collected during a calibration scan is performed on-the-fly on a set of processing nodes, which in turn send summary data for every detector chip to an aggregator node. The analysis of the data is performed both on the processing and aggregator nodes, depending on the scan. Both data decoding and analysis are performed by a set of calibration \textit{workflows} which are part of the ALICE O$^{2}$ system.
For noise scans, the ALPIDE raw data are decoded on the processing nodes and the pixel hits are sent to the aggregator node, which acts as a hit counter. The hit counting per pixel is parallelized, while once all hits have been recorded, a second serial process normalizes the per-pixel hits to the total number of recorded events to extract the per-pixel fake-hit rate. In the noise scan, a single EPN receives a full detector event, meaning that in a single readout frame of a random EPN, there are always pixel hits from the full detector.
In all the other calibration scans, the data distribution to the EPNs guarantees that the data of a certain detector link always reach the same EPN during the full scan. This is fundamental since a single EPN needs all pixel hits of a detector link in order to extract the calibration parameters (e.g. to extract the mean threshold of a chip, all the chip pixels' thresholds are needed on a single EPN). 
In this case, the single processing nodes decode the raw data, count the hits, and extract the calibration parameters (e.g. pixel threshold through the derivative of the \textit{S-curve}). Having multiple processing instances is of crucial importance to speed up the data processing: every workflow is connected to the same array of pixel hits but it counts hits only for a specific subset of chips (i.e. if $M$ workflows are defined, the workflow $n$ counts the hits coming from chips with IDs satisfying the following rule: $ID \% M = n$). The parallelization of the analysis is provided at the level of a single processing node and by employing multiple processing nodes.
In all cases, the aggregator node, at the end of the scan, sends the aggregated summary data from all detector chips to the Condition and Calibration Database (CCDB). This data can be, for example, the list of noisy pixels, the mean threshold for each chip, the list of noisy double-columns, the \texttt{VCASN} values of every chip needed to tune the pixel thresholds, etc. For the scans needed to calibrate the detector (e.g. \texttt{VCASN scan}), the CCDB object is used at a later stage for the preparation of a new detector configuration. This step is of crucial importance to calibrate the sensors to the desired threshold and to mask the noisy pixels. Finally, the EPN farm is connected to a 100 PB disk buffer (EOS Open Storage) where debug data are stored at the end of calibration scans, with the exception of the noise scan.

Generally, 40 processing nodes are used to process the ITS2 calibration data, with the exception of the full threshold scan, where 80 nodes are used instead. Each processing node has 64 CPU cores, and during the calibration runs, the ITS2 is the only user\footnote{As an example, the short threshold scan uses 40 EPNs, 6 parallel processes to decode data and 5 parallel processes to count hits and extract the thresholds. Each process is executed with 1 thread. In this case, the CPU usage of each process is up to about 120\% and 65\% for the decoding and the threshold extraction, respectively. While the memory used by each process ranges between 5 and 6 GB independently of the process type.}.
The logic behind the choice of the number of EPNs to be used is to make the data processing and analysis aligned with the duration of the scan in order to minimize any additional overhead: the usage of the buffer would be an issue if the throughput would not be matched for the different elements in the chain. With the exception of the noise scan, the bottleneck of the data processing resides in the extraction of the calibration parameters from every single pixel (e.g. extraction of the pixel threshold and temporal noise). Due to this and to its time duration, the full threshold scan requires more EPNs as mentioned above. A larger number of EPNs reduces the TF rate (or hit rate) on each node, making the data processing lighter.
In the noise scan, the bottleneck is instead on the aggregator node receiving the single hits from multiple processing nodes. 
With 40 EPNs, a full re-calibration (tuning of the thresholds and masking of the noisy pixels) of the detector is possible within the time between two LHC fills (approximately 45 minutes) in case of issues. The other advantage of having on-the-fly processing is that the data quality assessment through the Quality Control (QC) software can be done immediately after the scan ends.

The QC software, which is a part of the ALICE O$^{2}$, consists of a set of tasks running on both FLPs and EPNs. The tasks on FLPs check the data integrity and calculate the fake-hit rate\footnote{The chip fake-hit rates are calculated on FLPs by a QC task and not taken from the EPN aggregator node during noise scans. This is so the QC task used in normal data taking can be reused, avoiding the implementation of a specific one for the noise scan running on the EPN aggregator node.} (only in noise scans), while the tasks on the EPNs are in charge of collecting summary calibration parameters for each chip, e.g., mean thresholds of chips. They are shown as yellow boxes in Fig.~\ref{fig:fullscheme}. A merger node aggregates the data from the different FLPs/EPNs and periodically stores the data in the CCDB during the scan. A QC graphical interface allows the user to display the collected data as histograms, which can be used to assess the quality of every calibration scan.

\section{Results}\label{sec:results}
In this section, results related to in-pixel thresholds and temporal noise, pixel analogue pulse shape, and detector fake-hit rate are discussed. 

%
\subsection{Threshold and temporal noise}\label{sec:thrnoise}
\begin{figure}[h!]
    \centering
    \includegraphics[width=1.0\textwidth]{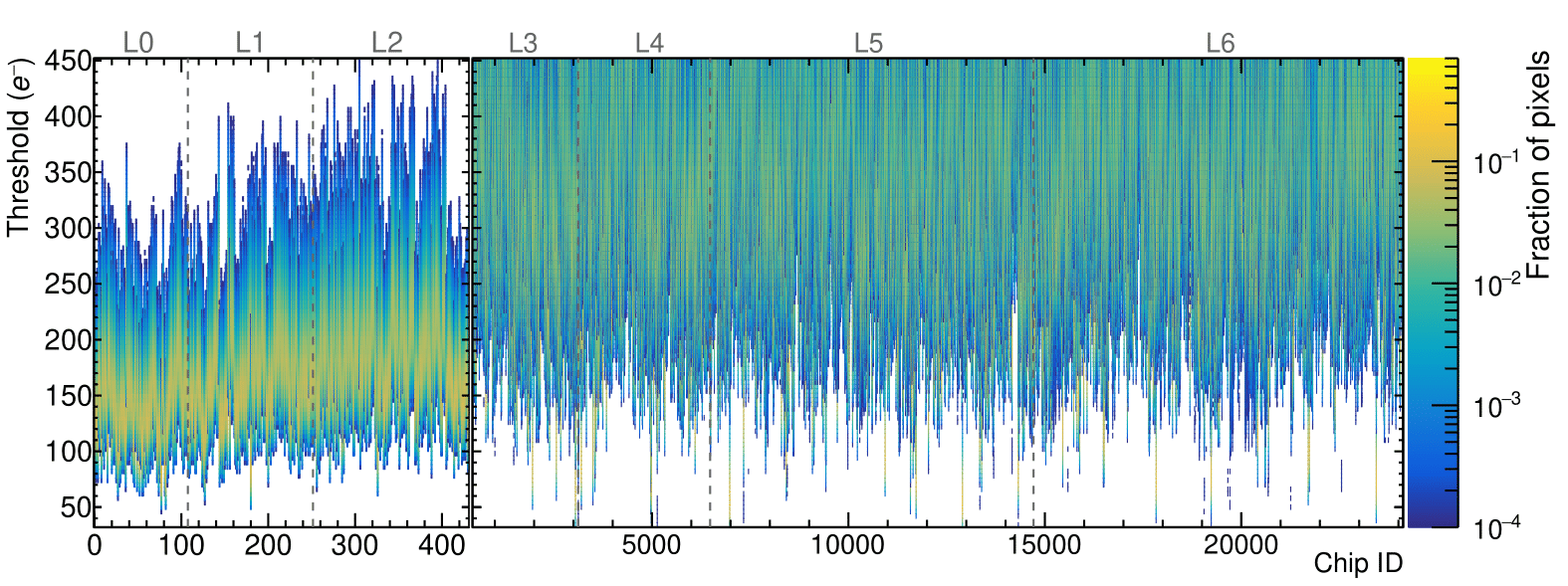}
    \includegraphics[width=1.0\textwidth]{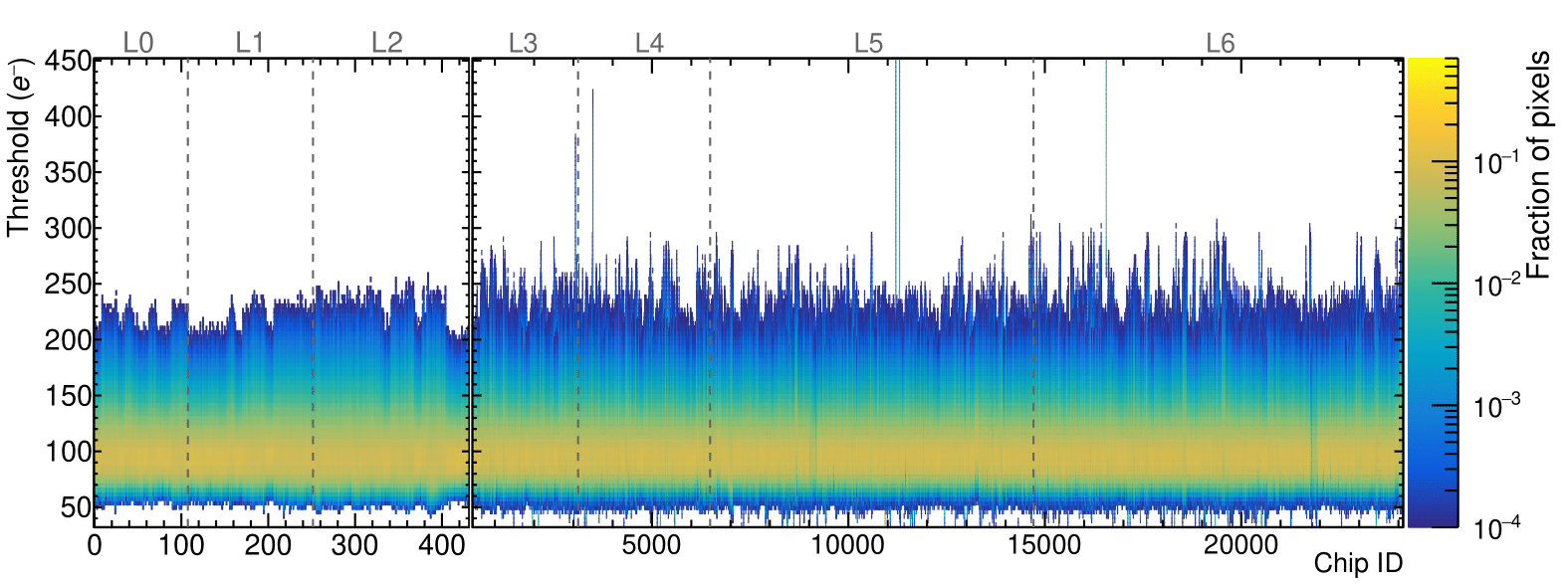}
    \caption{Pixel threshold distributions for every chip of ITS2 from two different full threshold scans, both recorded in June 2025. The top figure shows the untuned case where default settings for VCASN and ITHR (50 DACs) have been adopted. The bottom panel refers to the case where a threshold tuning to 100 $e^-$ has been performed. The $x$ axis of both plots is split into two parts: IB chips on the left, and OB ones on the right. The $y$ axis maximum is set to 450 $e^-$ since above this limit the threshold cannot be reliably extracted, given that the maximum injected change is 500 $e^-$. See text for more details on outlier chips. }
    \label{fig:thrchip}
\end{figure}

\begin{figure}
    \centering
    \includegraphics[width=0.5\linewidth]{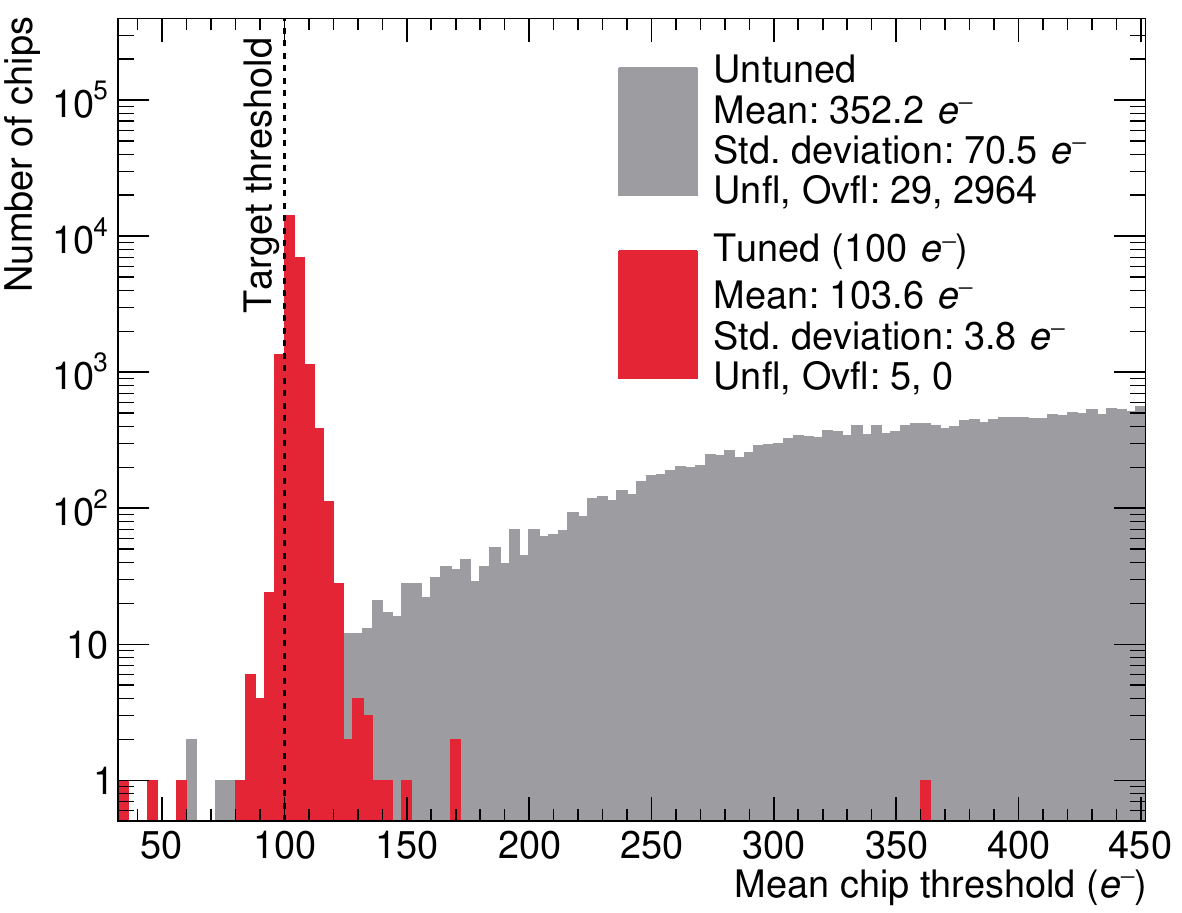}
    \caption{Distributions of the mean chip thresholds in both the tuned and untuned cases. The vertical dashed line represents the target threshold of 100 $e^-$. The legend reports mean, standard deviation, and the number of entries in the underflow and overflow bins. See text for more details on outlier chips. }
    \label{fig:chipavg}
\end{figure}

\begin{figure}[h!]
    \centering
    \includegraphics[width=1.0\textwidth]{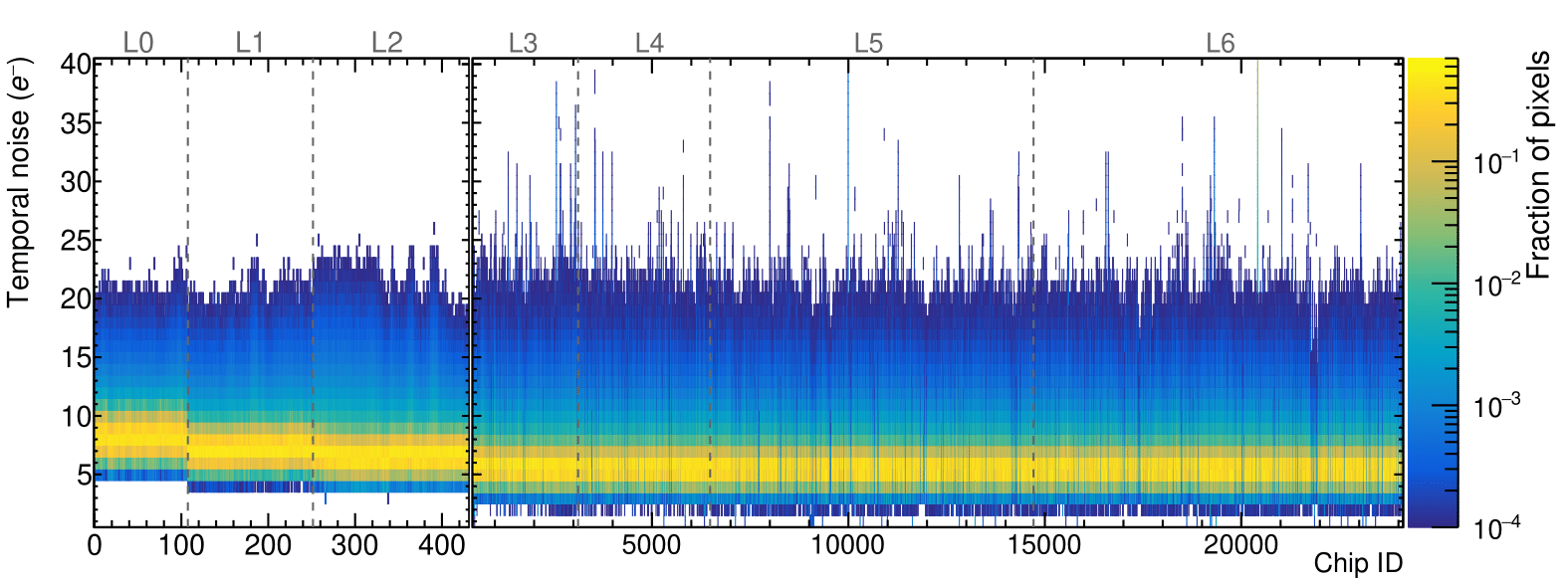}
    \includegraphics[width=1.0\textwidth]{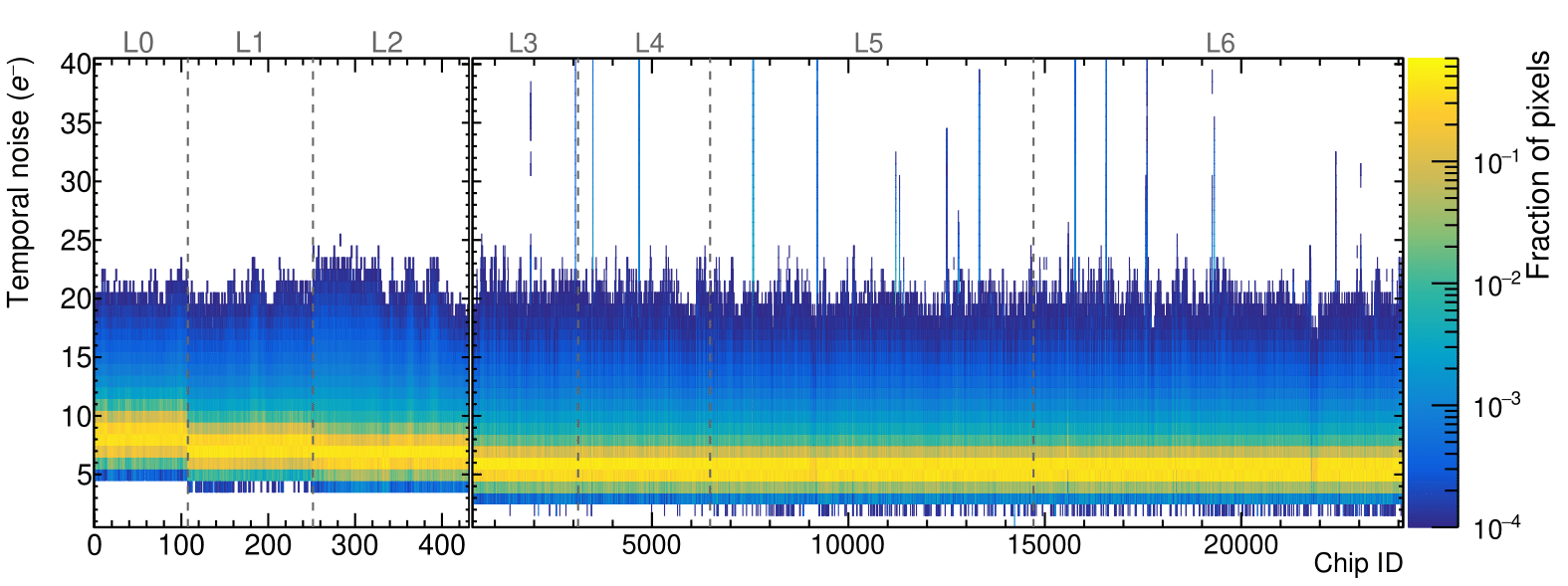}
    \caption{Pixel temporal noise distributions for every chip of ITS2 from two different full threshold scans, both recorded in June 2025. The top figure shows the untuned case where default settings for VCASN and ITHR (50 DACs) have been adopted. The bottom panel refers to the case where a threshold tuning to 100 $e^-$ has been performed. The $x$ axis of both plots is split into two parts: IB chips on the left (chip ID from 0 to 431) and OB ones on the right (chip ID from 432 to 24119). See text for more details on outlier chips.}
    \label{fig:noisechip}
\end{figure}

\begin{figure}[h!]
    \centering
    \includegraphics[width=1\textwidth]{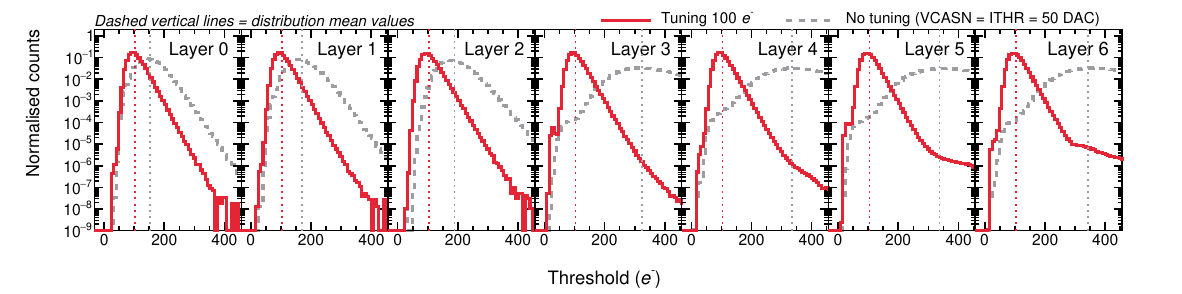}
    \includegraphics[width=1\textwidth]{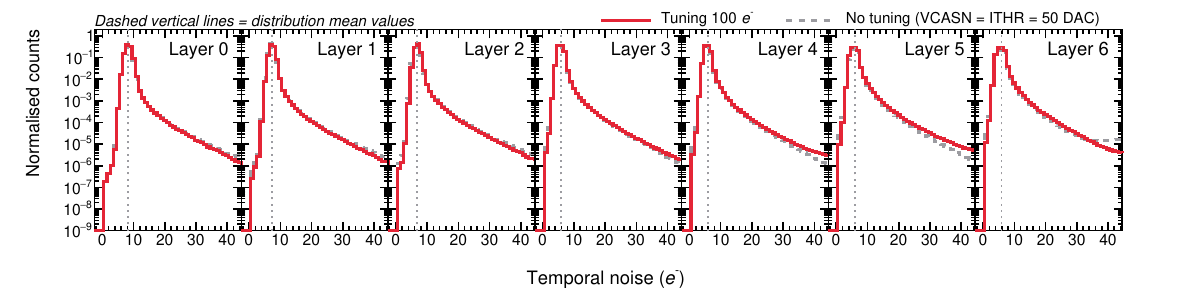}
    \caption{Distributions of the pixel threshold (top) and temporal noise (bottom) for each ITS2 layer from two different full threshold scans, both recorded in June 2025. The red distributions refer to the case with thresholds tuned to 100 $e^-$ while the gray ones to the untuned case with the default settings for \texttt{ITHR} and \texttt{VCASN} (50 DAC for both). Distributions are normalized to their integral in order to have a direct comparison between the layers. The vertical dashed lines represent the mean of the distributions.}
    \label{fig:1ddist_thrnoise}
\end{figure}

Figure~\ref{fig:thrchip} shows the threshold distributions for each of the 24120 chips of ITS2 measured in two different full threshold scans, both recorded at the end of June 2025. The top panel of Fig.~\ref{fig:thrchip} refers to the untuned scenario where both ITHR and VCASN DAC registers are set to 50 DAC units, the respective default values for all chips. The bottom panel of Fig.~\ref{fig:thrchip} instead depicts the threshold distributions after tuning the thresholds to 100 $e^-$. In the untuned case, it is possible to notice that IB chips had, on average, a different threshold compared with OB even if the DAC settings were the same for all chips. This is due to the respective accumulated doses up to June 2025 (see also Sec.~\ref{sec:thrstab} for more details on threshold variations). For an estimate of the expected radiation levels, see Table 1.2 in Ref.~\cite{ALICE:2013nwm}. The effect is fully compensated by the threshold tuning procedure.
By comparing the top and bottom panels of Fig.~\ref{fig:thrchip}, it is possible to observe that the calibration framework can precisely tune all chips to the desired average threshold. The remaining threshold dispersion after the tuning is due to the pixel-by-pixel variations within a chip, which amounts to about 20 $e^-$ (standard deviation). However, after the tuning, there are still five chips that have a threshold distribution containing several pixels with higher thresholds: the four outliers in  Fig.~\ref{fig:thrchip} bottom, with chip~${\rm ID} = 3051$, 3496, 11206, 11301, correspond to chips having a mean threshold between 140 and 170 $e^-$ (depending on the chip) and a long tail up to 450 $e^-$. Another outlier, having chip~${\rm ID} = 16563$, corresponds to a chip with a mean threshold of about 360 $e^-$ without a long tail. On the opposite side, there are also mainly four chips (chip ${\rm ID} = 1960$, 10933, 20510, 23435) that have very low thresholds for the vast majority of the pixels: they are visible below 50 $e^-$ in Fig.~\ref{fig:thrchip} bottom. These nine chips all belong to OB layers and are known to be problematic.\\ \\
The mean chip threshold distribution, as derived from the same dataset presented in Fig.~\ref{fig:thrchip}, is shown in Fig.~\ref{fig:chipavg} in both the tuned and untuned cases. The chip-by-chip standard deviation is only 3.8 $e^-$ after the threshold tuning, while the mean threshold is 103.6 $e^-$. The latter is slightly above the target threshold of 100~$e^-$. This is again due to radiation accumulation since the thresholds were tuned 3 months before. The mean chip thresholds are calculated by excluding non-working pixels (also called \textit{bad} pixels) while the fully non-working chips are excluded from the histograms (see Sec.~\ref{subsec:badpix}).
\\ \\ 
Similarly, Fig.~\ref{fig:noisechip} shows the temporal noise calculated from the same two threshold scans described above. A mean temporal noise ranging approximately from 5 to 8 $e^-$ is measured for all detector chips in both cases, meaning that the threshold tuning does not have a significant influence on the temporal noise apart from a small fraction of pixels in the upper tail of the distributions for which the noise is reduced when applying the tuning. In addition, several chips in the OB, only in the untuned case, have distributions filled with a very few pixels (vertical blue lines in Fig.~\ref{fig:noisechip}, top panel, interrupting the yellow band).
These chips have a mean threshold above 400--450 $e^{-}$ with default settings of \texttt{VCASN} and \texttt{ITHR} (untuned case). As a consequence, their pixel thresholds and the temporal noise cannot be reliably calculated for a large fraction of pixels (see also Fig.~\ref{fig:thrchip}). This is fixed by the tuning procedure as it is visible in the bottom panel of Fig.~\ref{fig:noisechip}. As previously discussed, there are also nine chips that were found to exhibit problematic threshold values. They are also outliers in terms of temporal noise, as it is visible especially in the bottom panel of Fig.~\ref{fig:noisechip}. Moreover, as for the thresholds, there is another group of a few chips in the OB with long tails towards 40~$e^-$. These are again peculiar chips that, however, have a good threshold distribution after the tuning. Finally, the chips of the IB show a trend in the temporal noise: chips from 0 to 107 (layer 0) show a mean noise of about 8 $e^-$, which decreases to about 6 $e^-$ when moving to layer 2 (up to chip 431). This is due to the different accumulated radiation doses of the layers. On the contrary, OB chips uniformly show a noise of about 5~$e^-$, comparable to the $\sim$4~$e^-$ measured in single-chip tests in the laboratory~\cite{AglieriRinella:2017lym}.
\\ \\
Figure~\ref{fig:1ddist_thrnoise} shows the pixel threshold and noise distributions for each detector layer, before and after the threshold tuning, as derived from the same dataset presented in Fig.~\ref{fig:thrchip}. Figure~\ref{fig:1ddist_thrnoise} represents a more detailed comparison between the tuned and untuned cases. Also in this case, bad pixels are removed from the distributions (see Sec.~\ref{subsec:badpix}). 
It is possible to see that the threshold tuning shifts the mean threshold towards the desired target of 100 $e^-$, reducing the tail at large thresholds and hence the standard deviation of the distribution (e.g., for layer 0 the standard deviation decreases by almost a factor 2 after a tuning while keeping its ratio to the distribution mean almost constant within 5\%). In the threshold distributions of layers 3, 5, and 6, it is possible to note a certain number of entries at very low thresholds which come from the outlier chips discussed before in Fig.~\ref{fig:thrchip}. On the contrary, the noise distributions remain very similar in the two scans as previously described. Also, for the temporal noise, discrepancies in the right tails of layers 5 and 6 are due to the outlier chips previously discussed. In general, as also observed in single ALPIDE laboratory tests, the temporal noise distributions deviate from a Gaussian shape, showing a pronounced tail toward higher values. Nevertheless, the average threshold-to-noise ratio remains around 20 for tuned thresholds of 100 $e^{-}$, that is within the requirements for standard detector operations. \\
%
\subsection{Bad pixels and non-working chips}\label{subsec:badpix}
The threshold scan of all detector pixels with tuned thresholds allows us to estimate the number of malfunctioning pixels, also referred to as bad pixels for simplicity. In the threshold scan, a pixel is considered bad if the threshold cannot be reliably extracted from the S-curve. A good S-curve must have at least one charge point with 0 hits and another one with the maximum number of hits expected (hit probability equal to 1). The charge for the point with 0 hits must also be lower than that of the point with the maximum number of hits. An example of a good S-curve is shown in Fig.~\ref{fig:scurve}. If these conditions are not met, the pixel is considered bad. This can be due to various root causes:
\begin{enumerate}
    \item an excessive noise or a too-low threshold leading to hits independent of the thresholds;
    \item a too-high threshold preventing the pixel from firing with 100\% probability within the charge range;
    \item a malfunction in the charge injection circuit, preventing the pixel from being stimulated;
    \item a malfunction in the pixel or its respective readout circuitry.
\end{enumerate}
The different root causes can be partially determined through other tests (see also Sec.~\ref{sec:noisydcols}).
\begin{figure}[h!]
    \centering
    \includegraphics[width=1.0\linewidth]{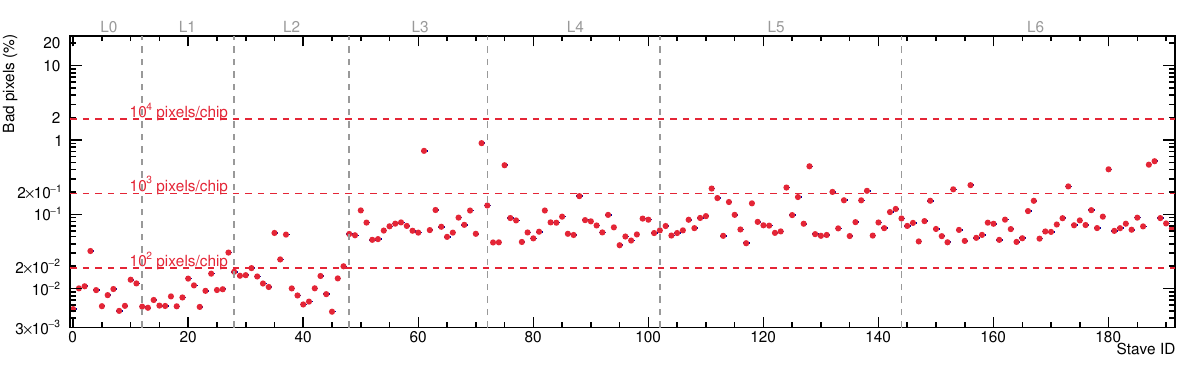}
    \caption{Percentage of bad pixels for every stave of ITS2 as extracted from a full threshold scan with thresholds tuned to 100 $e^-$. The fully non-working chips are excluded from the calculation. The number of chips per stave is 9, 112 and 196 for IB, ML, and OL staves, respectively. Vertical dashed lines separate the different layers, while the horizontal red lines are used as a reference to indicate the percentages corresponding to 100, 1000, and 10000 bad pixels per chip. See text for more details.}
    \label{fig:deadpix}
\end{figure}
Figure~\ref{fig:deadpix} shows the percentage of bad pixels in every detector stave. Fully non-functional chips are excluded from this calculation. These amount to 105 chips, distributed across different OB staves, that either have readout issues or were damaged during the stave assembly phase. It is possible to note the clear separation between IB and OB staves: by construction, IB staves were assembled with higher-quality chips, which explains the observed difference. 
In total, ITS2 has about 1.32 $\times$ 10$^7$ bad pixels, which corresponds to about 0.10\% of the total number of pixels. The fully non-working chips instead contribute to 0.43\% of the total number of pixels. \\  

\subsection{Masking of noisy pixels: digital and noise scans}\label{sec:noisydcols}
\begin{figure}[h!]
    \centering
    \includegraphics[width=1.0\linewidth]{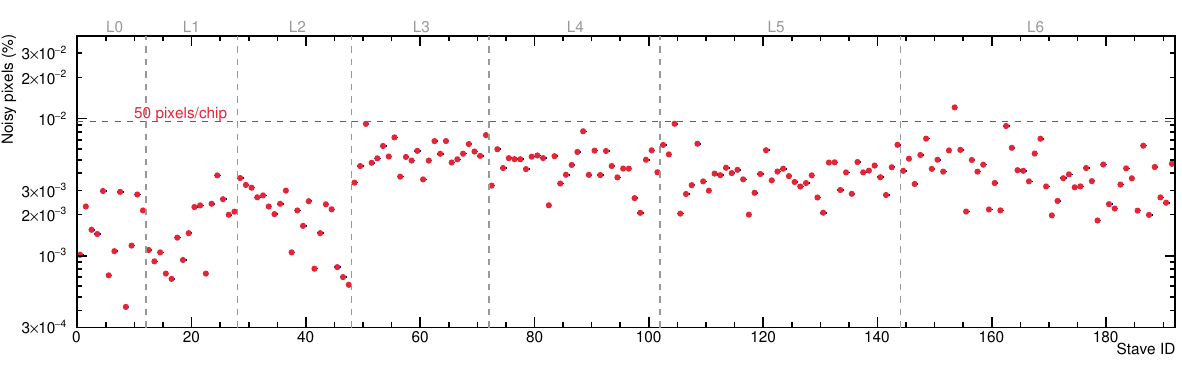}
    \caption{Percentage of noisy pixels for every detector stave in a noise scan performed in September 2023 after a threshold tuning to 100 $e^-$. The number of chips per stave is 9, 112 and 196 for IB, ML, and OL staves, respectively. Vertical dashed lines separate the different layers, while the horizontal red line is used as a reference to indicate the percentage corresponding to 50 noisy pixels per chip.}
    \label{fig:noisypix}
\end{figure}

A set of ALPIDE double columns are masked during data acquisition based on the results from a digital scan of the full detector. This is a measure to minimize the risk of occasionally stalling the readout. The IB does not have problematic double columns, while for the OB staves a maximum of about 0.03\% of the total number of double columns in a stave is masked. This corresponds to about 17(30) double columns in a full ML(OL) stave, which results in a negligible fraction. \\ \\
Once the thresholds are tuned to 100 $e^-$, the noise scan allows us to determine the noisy pixels in the detector and mask them. Using the definition of noisy pixels outlined in Sec.~\ref{sec:noise-calib}, Figure~\ref{fig:noisypix} shows the percentage of noisy pixels for every detector stave. It is possible to see that the overall percentage always stays well below (or close to, for very few staves) 0.01\%, meaning less than 50 pixels per chip are masked on average on top of the noisy double columns previously discussed.
During the full threshold scan, both the noisy pixels and the noisy double columns previously mentioned are masked. As a result, the bad pixel percentage shown in Fig.~\ref{fig:deadpix} already includes the contribution from noisy pixels and noisy double columns, which anyway represent a small fraction of the total number of bad pixels.
\\ \\
Overall, considering the number of bad pixels obtained from the full threshold scan, which includes noisy pixels and double columns, the maximum percentage of malfunctioning pixels for the whole detector is about 0.10\%. In addition, 0.43\% of the pixels belong to non-working chips. The detector did not show any significant deviations from these numbers since June 2022 when the final calibration of the detector was achieved. In particular, no significant increase due to radiation was observed up to the end of November 2024.
\subsection{Effect of VRESETD on the pixel thresholds}\label{subsec:resultsVRESETD}
\begin{figure}[h!]
    \centering
    \includegraphics[width=0.7\linewidth]{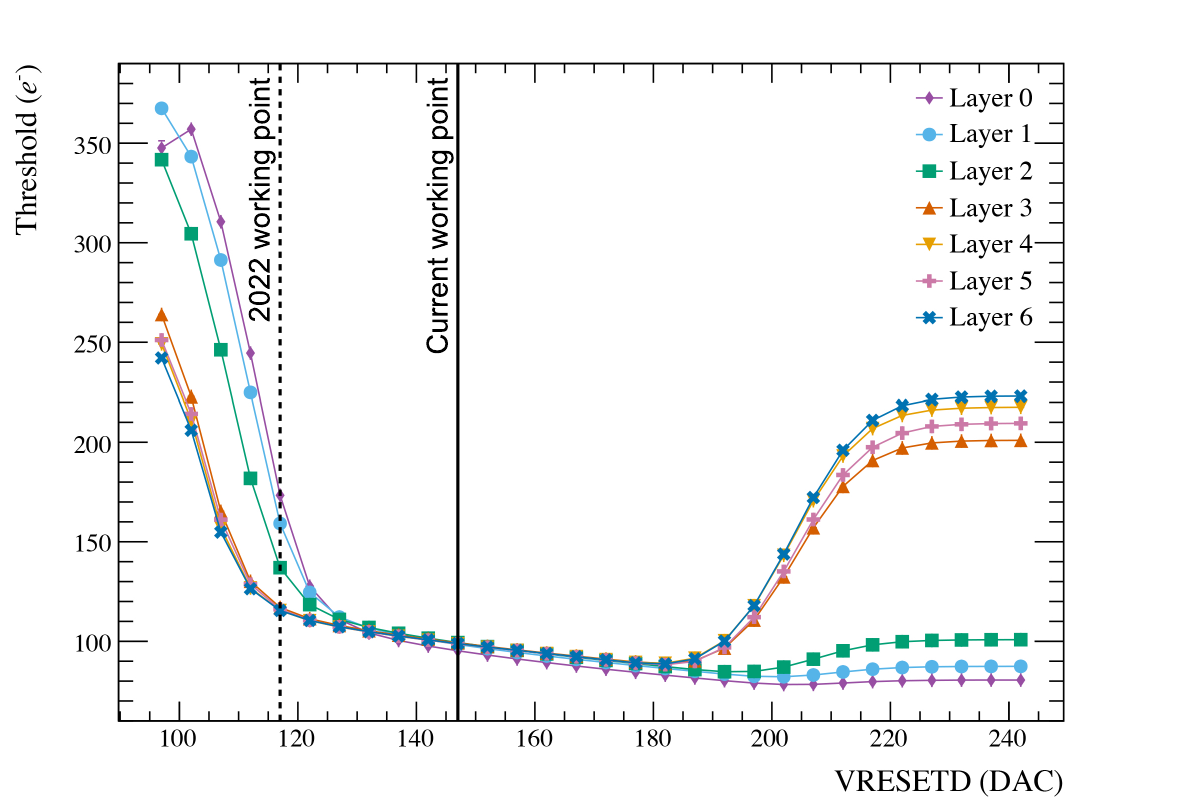}
    \caption{Results from the \texttt{VRESETD two-dimensional scan} recorded in July 2023. The $x$ axis represents the value at which the \texttt{VRESETD} DAC is set, the $y$ axis represents the mean threshold per layer in $e^-$. The error bars on the mean thresholds are estimated as the standard deviation of the mean chip threshold distribution per layer at a given \texttt{VRESETD} divided by the square root of the number of entries in the distribution. The bars are smaller than the marker size. The solid black line indicates the DAC setting in use since 2023 during standard operations. The dashed black line indicates the DAC setting used until 2022.}
    \label{fig:vresetd}
\end{figure}

The \texttt{VRESETD} scan (see Sec.~\ref{sec:vresetd-scan}) is used to check the correct operational reset voltage range depending on the desired threshold.
In Fig.~\ref{fig:vresetd}, the mean threshold for each layer is shown as a function of the \texttt{VRESETD} DAC setting. The pixel thresholds were tuned to a mean of 100 $e^-$ for a \texttt{VRESETD} of 147~DAC units. On the right and left side of the curves, where thresholds rapidly increase, layers behave differently with respect to each other. This is due to the different amounts of radiation accumulated: both the Total Ionizing Dose (TID) and the Non-Ionizing-Energy Loss (NIEL) accumulated on the IB layers are more than one order of magnitude higher than the amounts accumulated on the OB layers. For a more detailed estimate of the expected radiation levels, see Table 1.2 in Ref. \cite{ALICE:2013nwm}. Transistor properties are altered by the TID, while pixel leakage current increases with the NIEL, both impacting the working point.
Until the end of 2022, the default \texttt{VRESETD} DAC setting for all chips was 117 DAC units. 
In 2023, it was decided to set the new \texttt{VRESETD} working point to 147 DAC units. As it is visible in Fig.~\ref{fig:vresetd}, the older set point of 117 DAC counts results in an increase of the thresholds of up to 80\% in the innermost layers compared to the 100~$e^-$ measured at 147 DAC units. In addition to an increase in the threshold, the threshold spread increases due to the strong \texttt{VRESETD} dependence combined with pixel-by-pixel variations.
With \texttt{VRESETD} set to 147 DAC counts, all layers exhibit a small dependence of the charge threshold on \texttt{VRESETD} and hence a good uniformity in agreement with previous studies during the production and commissioning phases of the detector. 
\subsection{Analogue pulse shape}
\begin{figure}[h!]
    \centering
    \includegraphics[width=0.6\textwidth]{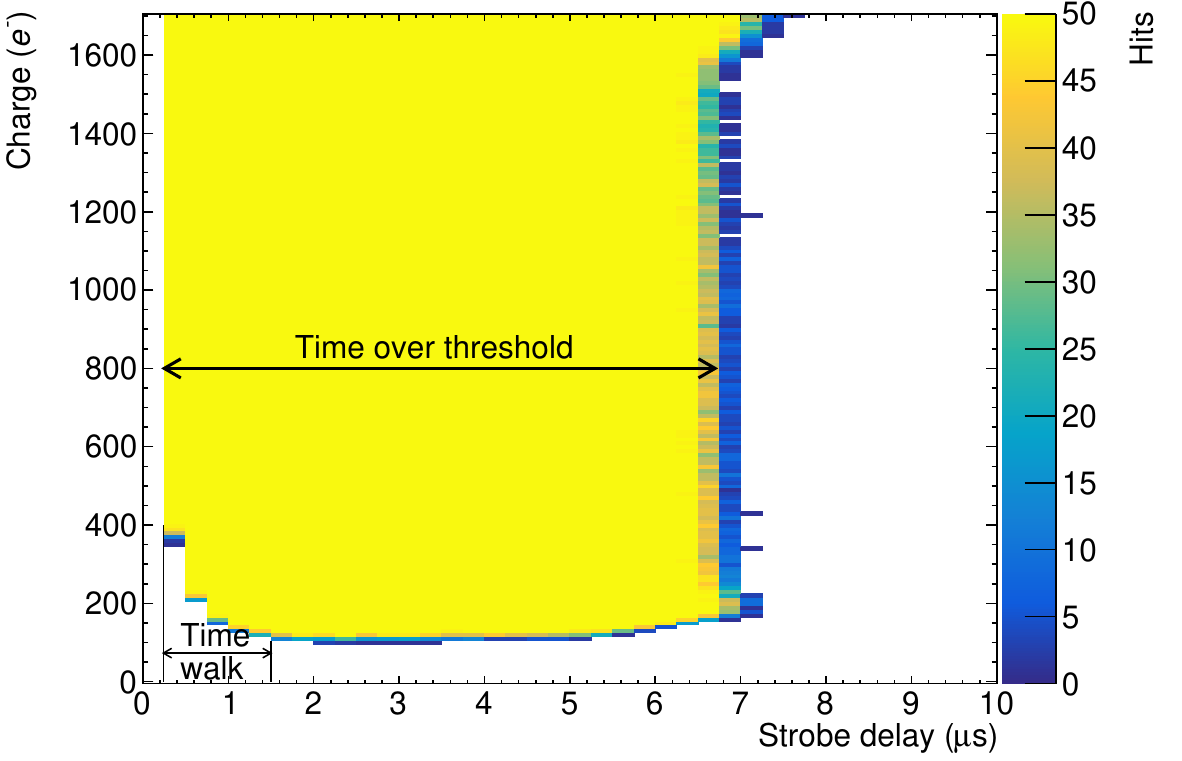}
    \caption{Two-dimensional histogram showing the injected charge as a function of the strobe delay for a pixel of ITS2. The colored scale refers to the number of hits recorded for each charge--delay pair. The ALPIDE is operated with the maximum signal clipping (\texttt{VCLIP} $=$ 0 DAC units) and the pixel under study has a tuned threshold of about 100 $e^-$. The double arrows indicate the time-over-threshold for a charge of 800 $e^-$ and the \textit{time walk}. See text for more details.}
    \label{fig:ps2d}
\end{figure}

Figure~\ref{fig:ps2d} shows the typical output of the two-dimensional pulse shape scan described in Sec.~\ref{sec:pulseshape}. The injected charge is shown as a function of the strobe delay for a single pixel of the detector with maximum clipping (\texttt{VCLIP} $=$ 0 DAC units). The colored axis represents the number of recorded hits for each charge--delay pair. No reverse substrate bias voltage is applied to the sensors, and the pixel threshold is tuned to 100 $e^-$. The features of Fig.~\ref{fig:ps2d} are summarised in the following:
\begin{itemize}
    \item the maximum number of 50 hits is obtained when the pixel analogue signal is above the threshold and the strobe is in coincidence with the digital output for each injection;
    \item the bottom part of the plot with no hits refers to the case of a signal always below threshold;
    \item for every charge the width of the region with more than 25 hits (the 50\% point as in the definition of the in-pixel threshold) is an estimation of the \textit{time-over-threshold} (ToT). This represents the time the analogue signal, which depends on the charge, spends above the threshold. During this time, the \texttt{STROBE} is in coincidence with the digital output;
    \item for each charge, the Time of Arrival (ToA) is estimated as the time gap between the point with a strobe delay equal to 0 and the first point with at least 25 hits (the 50\% point);
    \item the ToA depends on the injected charge, for charges below a few hundred electrons. The difference between the maximum and minimum ToA is defined as \textit{time walk}. This is of the order of 1 $\upmu$s;
    \item the slightly larger ToT at a charge of about 200 $e^-$ (small charges) and 1600 $e^-$ (large charges) are due to non-uniformities of the in-pixel circuitry.  
\end{itemize}

\begin{figure}[h!]
    \centering
    \includegraphics[width=0.49\linewidth]{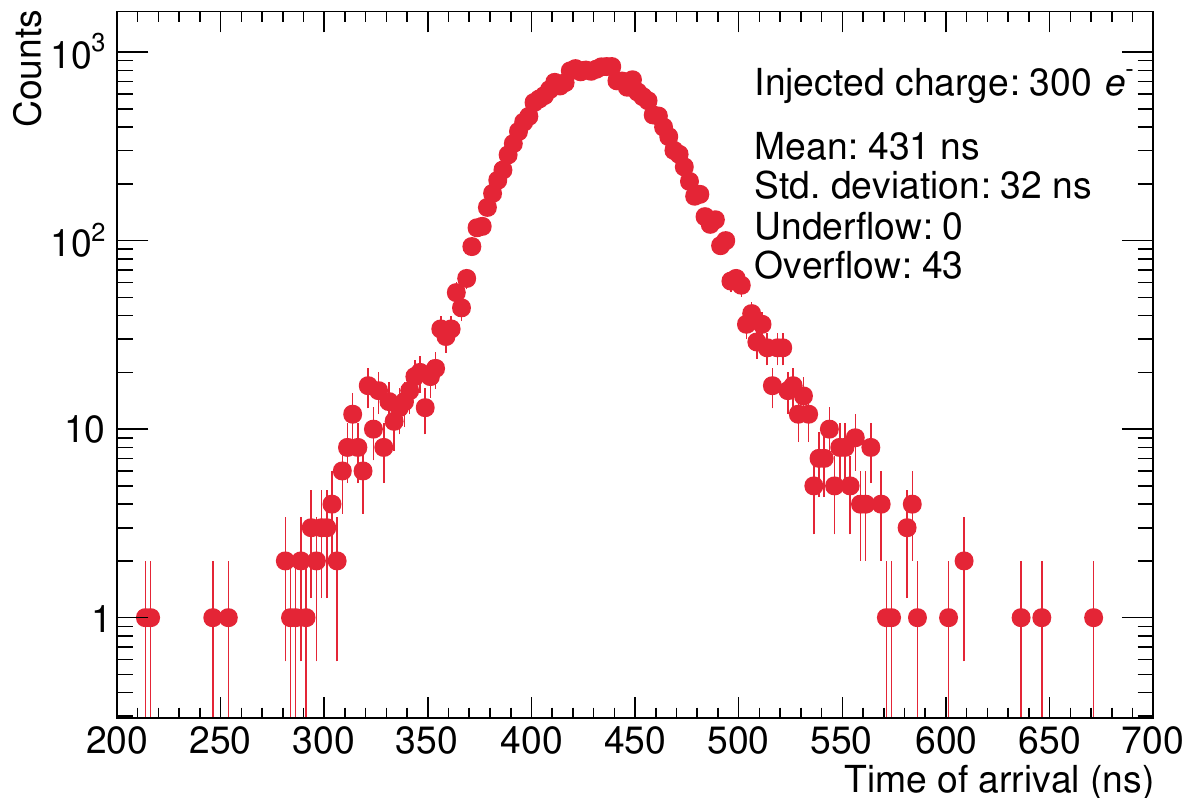}
    \hfill
    \includegraphics[width=0.49\linewidth]{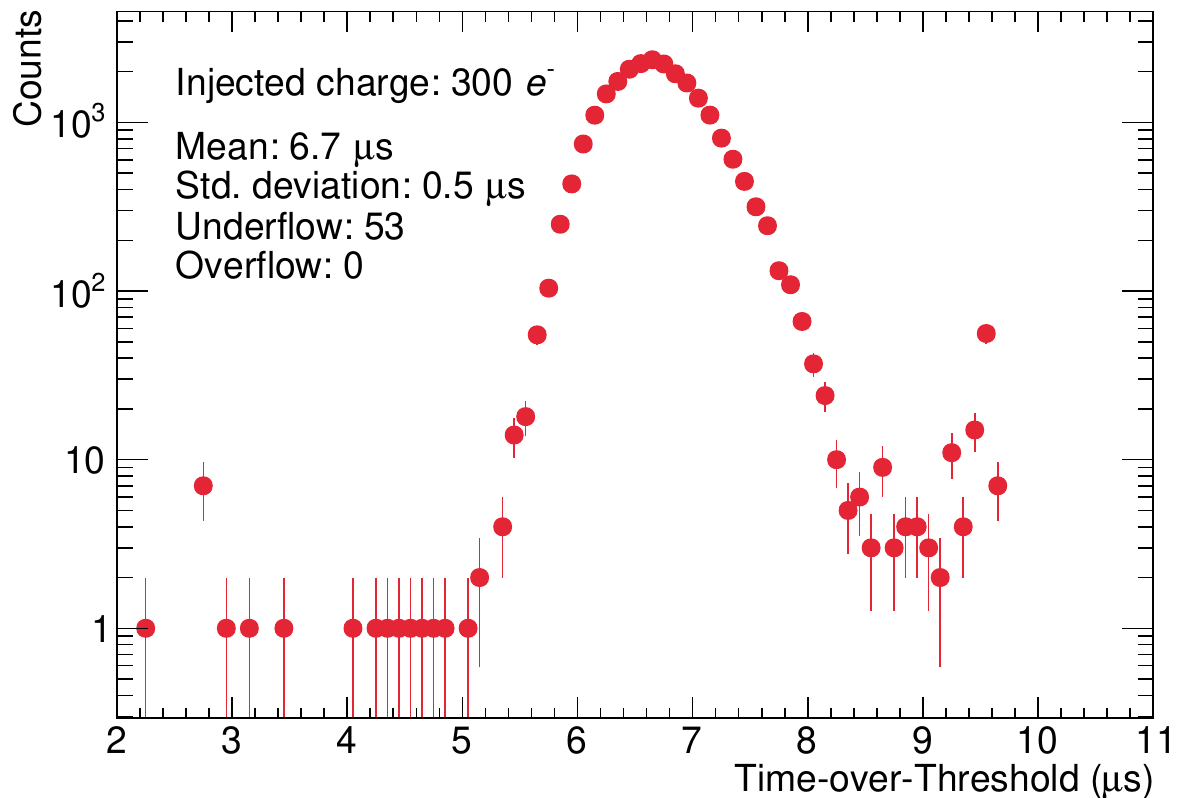}
    \caption{Left panel: distribution of the ToA of the ALPIDE analogue pulse calculated chip by chip as a mean of 1024 values coming from the pulse shape scan of 1 pixel row. Right panel: distribution of the ToT of the ALPIDE analogue pulse calculated chip by chip as a mean of 1024 values coming from the pulse shape scan of 1 pixel row. Thresholds were tuned to 100 $e^-$ for this measurement, and a charge of 300 $e^-$ was injected in the pixels. Error bars correspond to Poissonian errors of the counts.}
    \label{fig:ps}
\end{figure}

As described in Sec.~\ref{sec:pulseshape}, 1024 pixels per chip (1 row) are tested in the pulse-shape scan. Figure~\ref{fig:ps} shows the distribution of the pixel mean ToA (left panel) and of the mean ToT (right panel) calculated for every chip by injecting a test charge of 300 $e^-$ in each of them. The in-chip standard deviation is about 50~ns for the ToA and about 0.3 $\upmu$s for the ToT. As it is possible to see from Fig.~\ref{fig:ps}, the mean ToA\footnote{The usage of the ALPIDE internal sequencer for the generation of pulse and strobe signals, as explained in Sec.~\ref{sec:alpide}, ensures the precise estimation of the ToA, without the need of accounting for clock cycle offsets, extra delays between pulse and strobe, etc.} and ToT are 431 ns and 6.7 $\upmu$s, respectively, in agreement with what is expected from the design of the chip. The few entries in the overflow/underflow bin reported in the legends of the ToA and ToT histograms of Fig.~\ref{fig:ps} correspond to the chips having a ToA up to 6--7 $\upmu$s with a very small ($<$2 $\upmu$s) ToT. This can be due to a higher (compared to the tuned value of 100 $e^{-}$) threshold for the scanned pixels, an issue in the shaping/digitization of the signal inside the pixels, or an issue with the charge-injection circuitry.   
\subsection{Threshold stability over time}\label{sec:thrstab}
\begin{figure}[h!]
    \centering
    \includegraphics[width=0.98\linewidth]{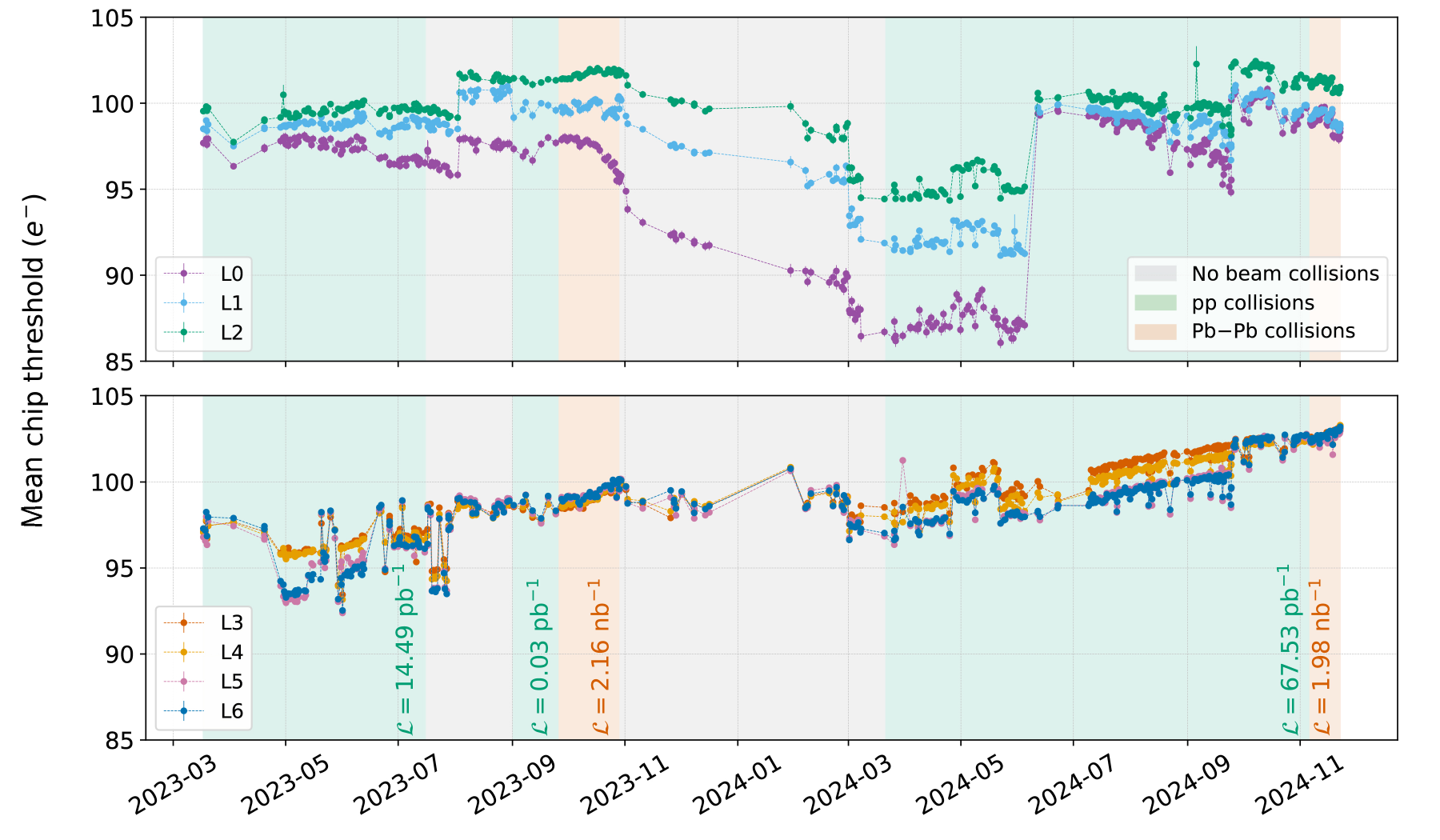}
    \caption{Evolution of the mean in-pixel thresholds per layer, from March 2023 to November 2024. All the threshold runs are recorded either during long (weeks) periods without beam or at the end of every LHC fill once the beam is dumped.
    The error bar for the threshold is calculated as the ratio between the standard deviation and the square root of the number of entries in the per-chip threshold distributions. For the majority of the data points, the error bar is smaller than the marker size. The gray bands indicate the long periods without beam collisions, the green and orange bands indicate periods with pp and Pb--Pb collisions, respectively. For each band, the value of the integrated luminosity delivered to ALICE during the corresponding period is reported.}
    \label{fig:thr_stability}
\end{figure}

Monitoring the mean thresholds over time is crucial to ensure the stability of the performance of the detector during data-taking operations.
Figure \ref{fig:thr_stability} shows the evolution of the threshold during a long period, from March 2023 to November 2024.
\begin{figure}[h!]
    \centering
    \includegraphics[width=0.98\linewidth]{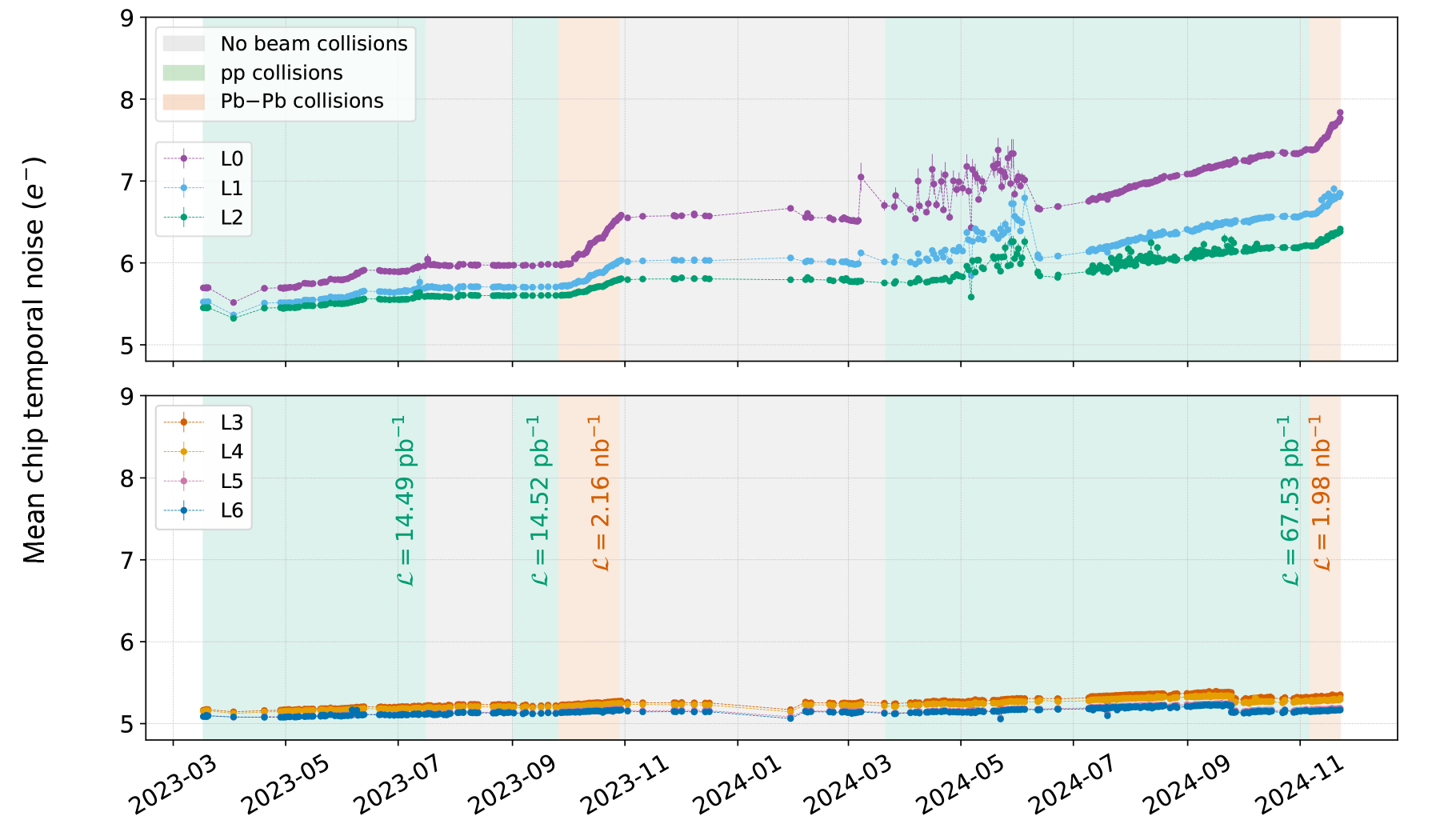}
    \caption{Evolution of the mean in-pixel temporal noise per layer, from March 2023 to November 2024. The temporal noise is extracted from threshold scan runs recorded either during long (weeks) periods without beam or at the end of every LHC fill once the beam is dumped. The error bar for the noise is calculated as the ratio between the standard deviation and the square root of the number of entries in the per-pixel noise distributions. The gray bands indicate long periods without beam collisions, the green and orange bands indicate periods with pp and Pb--Pb collisions, respectively. For each band, the value of the integrated luminosity delivered to ALICE during the corresponding period is reported.}
    \label{fig:enc_stability}
\end{figure}
During this period, the threshold remained in a range between 85 and 105 $e^-$. In order to achieve this, thresholds were re-tuned in June 2024 and October 2024, after a reduction to approximately 85 $e^{-}$ and 95 $e^{-}$ in the IB. Gray bands indicate long periods without beam collisions. During these periods, the detector is mostly powered off and occasionally turned on for tests or cosmic-ray runs. Green and orange bands indicate periods with pp and Pb--Pb collisions, respectively. For each band, the value of the luminosity delivered to ALICE integrated over the corresponding period is reported. Overall, the thresholds are stable over periods of months, but there are different kinds of variations described in the following.
After the second period without beam, a decrease of IB mean thresholds up to 15\% is observed. The decrease is due to the radiation load after the Pb--Pb data-taking period, which happened in October 2023. In fact, the threshold decrease for the layer 0 started already during the Pb--Pb collision period and continued, with a different slope, in the period without collisions. The same decreasing trend is visible after both the re-tuning points (June and October 2024). Starting from March 2024, a slight rise of the thresholds is visible in the OB layers: this effect can also be explained by radiation, and the different trend with respect to the IB is due to the different amounts of radiation accumulated over time.
From laboratory tests, it is known that the charge threshold of ALPIDE increases mildly up to a radiation dose of about 10 krad, after which a continuous decrease of the threshold begins.
This result underlines the importance of monitoring the mean thresholds over long periods in order to track potential inefficiencies for data-taking operations. It is demonstrated that a new threshold tuning can compensate for deviations accumulated over several months. 
\\ \\
Another effect induced by radiation is the increase of the temporal noise. Figure \ref{fig:enc_stability} shows the evolution of the temporal noise over the same period during which the threshold evolution was studied.
The increase of the temporal noise is greater for the layers of the IB, especially layer 0, and minimal for the layers of the OB. This indicates a clear correlation with the accumulated radiation dose. Another indication is that the noise remains stable during periods without beam, as indicated by the gray areas in the plot, and increases more steeply during periods with Pb--Pb collisions compared to those with pp collisions. Starting from March 2024, when thresholds in the Inner Barrel reached lower values as observed in Fig. \ref{fig:thr_stability}, the noise became unstable, with larger fluctuations from run to run. 
These fluctuations are attributed to the fact that the front-end settings are out of their optimal range due to radiation. This generates an instability of the high-speed links (increase in the number of bit errors), which leads to larger fluctuations of the temporal noise compared to other periods. The large error bars of the temporal noise in that period are due to larger chip-to-chip fluctuations with respect to other periods.
This instability, also observed in standard physics data taking, was solved with the re-tuning performed in June 2024. After that point, the noise continued to increase steadily with the accumulated radiation.
\\
\\
In addition to long-period threshold variations, small fluctuations of the thresholds over very short time periods (days) are visible in different time windows in Fig.~\ref{fig:thr_stability}, both for the IB and the OB. For example, several fluctuations are visible between May and August 2023 for OB, while for IB, a clear step is visible around August 2023. They are mainly due to the optimization of the analogue voltage drop correction to the chips.
\begin{figure}[h!]
    \centering
    \includegraphics[width=0.6\linewidth]{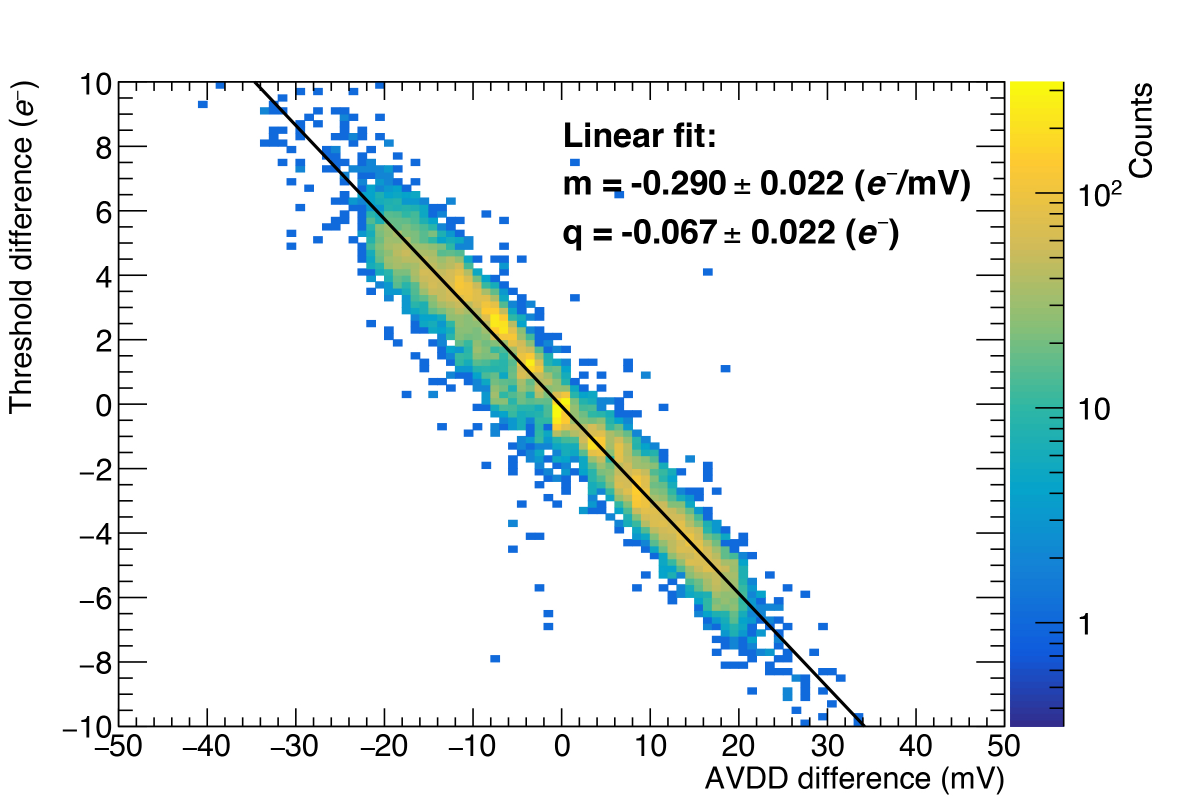}
    \caption{Threshold difference per HIC between a set of consecutive runs vs. the \texttt{AVDD} difference for the same runs. The black line represents the linear fit to the data where $m$ and $q$ are its slope and offset, respectively.}
    \label{fig:thr-avdd}
\end{figure}
\begin{figure}[h!]
    \centering
    \includegraphics[width=0.48\linewidth]{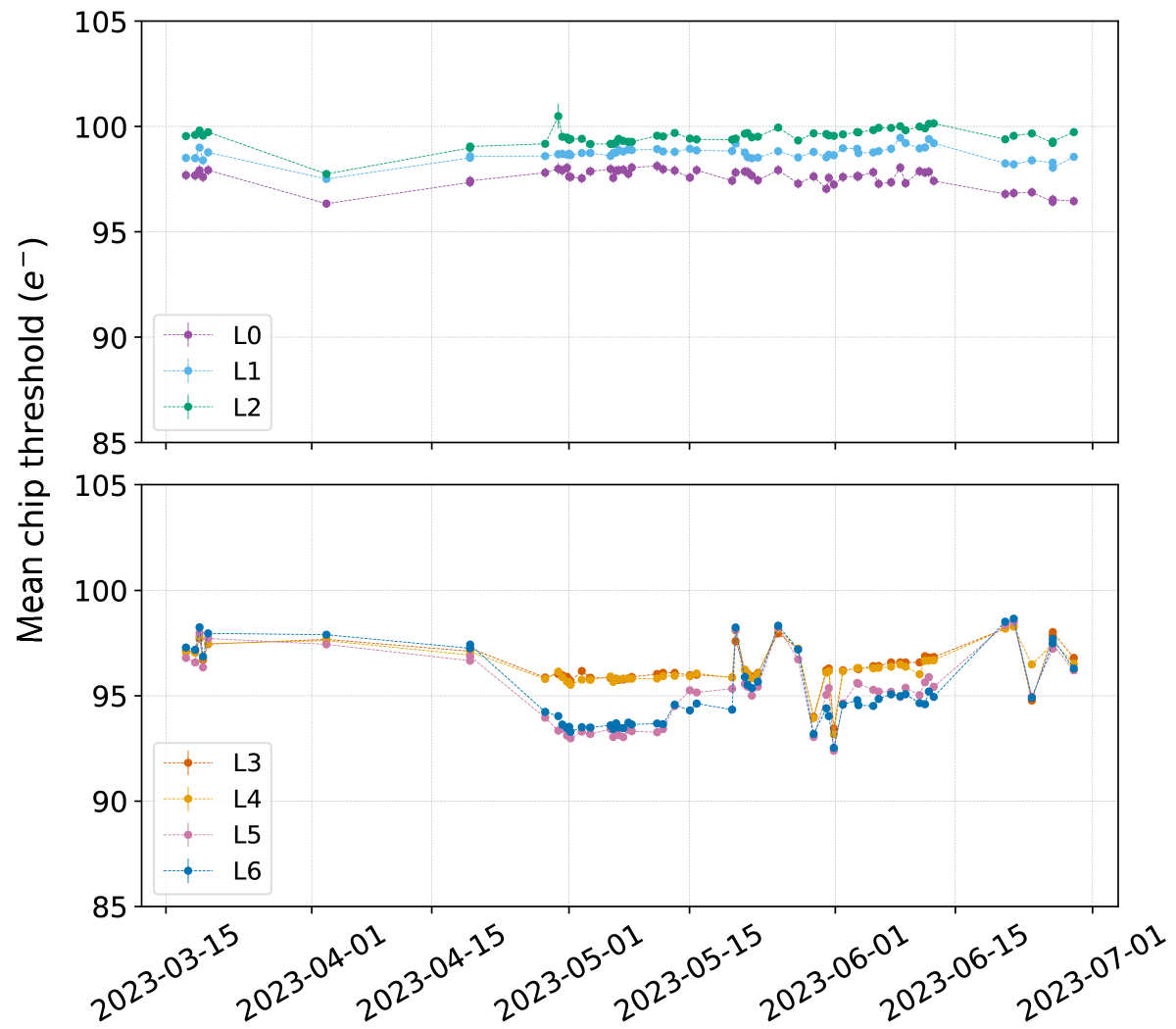}
    \includegraphics[width=0.48\linewidth]{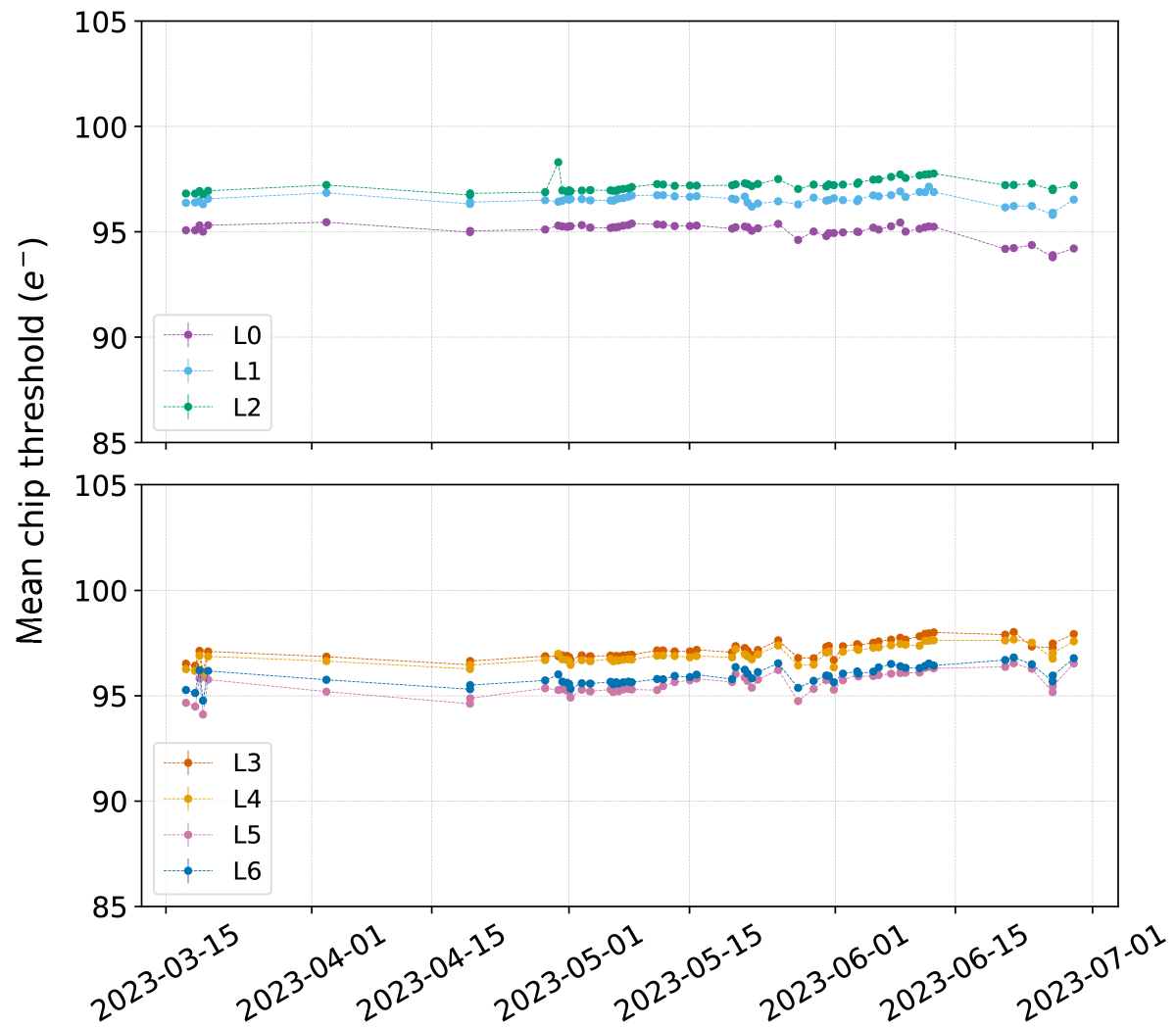}
    \caption{Threshold trend from March to July 2023 without (left) and with (right) correction of the HIC thresholds for changes in AVDD. The corrected thresholds were evaluated for ${\rm AVDD} = 1.8$ V. See text for more details. }
    \label{fig:thrcorrection}
\end{figure}
Figure \ref{fig:thr-avdd} shows a correlation plot of the threshold difference per HIC between a set of consecutive runs as a function of the \texttt{AVDD} difference per HIC for the same runs. A linear correlation between these quantities is visible. In particular, a positive change in \texttt{AVDD} corresponds to a negative change in the threshold, and vice versa. The magnitude of the threshold change is about 0.3 $e^-$ for a 1 mV change of \texttt{AVDD}.
Slope and offset found from the linear fit were used to correct the threshold of runs from March 2023 to July 2023 to a nominal \texttt{AVDD} value of 1.8 V. The threshold trend during these months before and after the correction is shown in Fig.~\ref{fig:thrcorrection}. It is clear that the measurement of the thresholds is very sensitive to \texttt{AVDD} variations: the fluctuations of 3--5~$e^-$ observed in Fig.~\ref{fig:thrcorrection} are due to fluctuations in \texttt{AVDD} of only 10--15~mV. 
The effect of the correction is visible especially for OB layers, and the remaining variations are at the level of 1 $e^{-}$. This suggests the presence of additional minor sources of fluctuations that are not explored in this study.
\subsection{Fake-hit rate stability over time}
\begin{figure}[h!]
    \centering
    \includegraphics[width=1.0\linewidth]{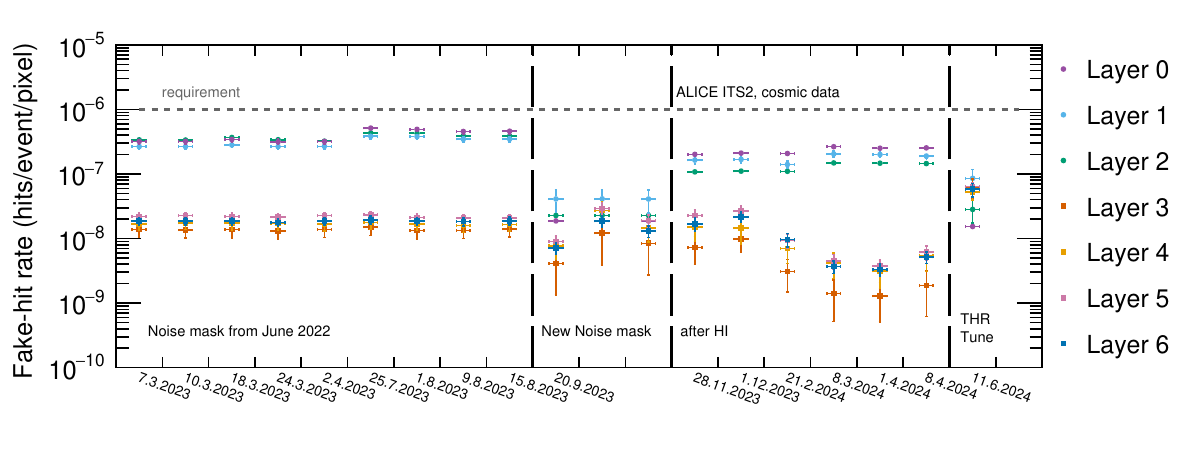}
    \caption{Trend of the fake-hit rate averaged for every detector layer. Cosmic runs in different periods are used to determine the fake-hit rate. The error bars represent half of the difference between the maximum and minimum fake-hit rate of the chips in a given layer. 
    See the text for more details on the four different periods separated by vertical dashed lines.}
    \label{fig:fhrtrend}
\end{figure}

Figure~\ref{fig:fhrtrend} shows the trend of the fake-hit rate for different time intervals corresponding to periods without beam when ALICE was taking data with cosmic rays. The fake-hit rate is averaged separately for each layer and expressed in hits/event/pixel. 
Starting from the left, the first region shows the trend for the first months of 2023 with a noise mask extracted in June 2022. 
The fake-hit rate remains stable below the design requirement (the gray dashed line in Fig.~\ref{fig:fhrtrend}) for the whole period after the noise scan, and after an integrated luminosity delivered to ALICE of 14.5~pb$^{-1}$ until mid-August 2023.
In the second region, the noise mask was updated before the start of the 2023 heavy-ion collision period. As a consequence, the fake-hit rate significantly dropped below 10$^{-7}$ hits/event/pixel.
In the third region, the fake-hit rate level increased by about an order of magnitude in IB layers and decreased gradually by approximately the same amount in the OB layers. This region refers to the period after the heavy-ion (HI, in the figure) collision campaign, where the LHC delivered a total luminosity of 2.16 nb$^{-1}$ to ALICE. This is mainly related to the decrease of the thresholds in the Inner Barrel layers and to their increase in the Outer Barrel layers, as shown in Fig.~\ref{fig:thr_stability}.
Finally, the fourth region shows that after the threshold tuning performed in June 2024, the fake-hit rate level stabilized between 10$^{-7}$ and 10$^{-8}$ hits/event/pixel. This highlights the correlation between the threshold and the fake-hit rate, showing the importance of maintaining the threshold at the target level to ensure the stability of the fake-hit rate.
Since multiple factors can influence the fake-hit rate of the detector, it is important to monitor it on large time scales. However, the fake-hit rate never exceeded the design limit of 10$^{-6}$ hits/event/pixel, allowing smooth data taking over the years.

\section{Summary}\label{sec:conclusions}
The new ALICE Inner Tracking System (ITS2) is made of 24120 silicon Monolithic Active Pixel Sensors (MAPS) called ALPIDE with a pixel pitch of 27$\times$29 $\upmu$m$^{2}$ and a total number of 12.6 $\times$ 10$^{9}$ pixels. The large number of channels and the power distribution to the detector, without active components close to the sensors, make the calibration of the sensors a challenging and mandatory step before recording any physics data. The calibration consists of tuning the in-pixel thresholds to a value at which the ALPIDEs are fully efficient while keeping the fake-hit rates as low as possible. Thresholds are generally tuned between 100 and 150 $e^-$ at which the ALPIDE chip is demonstrated to have a detection efficiency above 99\%. The fake-hit rate instead must stay below the design value of 10$^{-6}$ hits/event/pixel to limit combinatorics during track reconstruction.\\ \\
Several calibration procedures were developed to calibrate the chip temperature sensor, measure and tune the in-pixel thresholds, mask the problematic pixel columns, mask the noisy pixels, and measure the characteristics of the pixel signal response. In addition to that, a set of tools was also implemented to monitor the stability over time of thresholds and fake-hit rate.
The calibration data are processed on-the-fly on the ALICE computing farm. 
For standard physics data-taking, the thresholds are tuned to 100 $e^-$. The calibration framework developed for the ITS2 has been demonstrated to be able to precisely tune the thresholds of all the chips across the whole detector. The remaining threshold dispersion after the tuning is due to the pixel-by-pixel variations within a chip, which amount to about 20 $e^{-}$ (standard deviation). Instead, the standard deviation of the mean chip thresholds is only 3.8 $e^{-}$. It was observed that a re-tuning is needed about once a year since the mean thresholds can significantly decrease or increase depending on the radiation load as a function of the radial distance from the interaction point in the LHC. The maximum decrease, observed between March and June 2024, was about 15\%. Threshold fluctuations are also observed over short time periods and are attributed to variations of the ALPIDE analogue voltage. The observed variations had no significant impact on the data taking; however, it is of crucial importance to monitor the thresholds over time and, if needed, perform prompt retuning. \\ \\
Periods without beam collisions allow us to measure the residual fake-hit rate, which was observed to always be well below the design limit of 10$^{-6}$ hits/event/pixel. This is obtained by masking a negligible number of pixels in each chip: at maximum 50 (0.01\%) pixels per chip are masked. A new noise mask is generally produced when the fake-hit rate deviates significantly from the standard detector status. \\ \\
The calibration scans also allow us to determine non-working regions of the detector. 
A total of 0.53\% of pixels of ITS2 are not working. Among them, 81\% belong to non-working chips while the remaining ones are individual pixels not responding properly to charge injection, or with a firing probability greater than 10$^{-6}$ hits/event ($noisy$ pixels), or belonging to broken chip double columns (entirely masked during data taking).
During data taking, the typical time-over-threshold of the pixel analogue signals, with thresholds tuned to 100 $e^-$ on average, is 6--8 $\upmu$s with a time of arrival of about 430 ns. These values were extracted by injecting a charge of 300 $e^{-}$. The time of arrival is dependent on the charge for charges below 400~$e^{-}$. The difference between the maximum and minimum time of arrival, defined as \textit{time walk}, is of the order of 1 $\upmu$s. 
\\ \\
To conclude, the results confirm that the daily (for thresholds) or monthly (for noise) monitoring of the calibration parameters is an essential tool to ensure the efficient operation of the ITS2, and that a re-calibration of the detector is needed once per year for operations in LHC Run 3. The development of the calibration framework was done in a way that allows for a complete re-calibration of the detector between two LHC fills if needed.


\newenvironment{acknowledgement}{\relax}{\relax}
\begin{acknowledgement}
\section*{Acknowledgements}
F. Krizek acknowledges support by the Ministry of Education, Youth and Sports of the Czech Republic, project LM2023040.
\end{acknowledgement}

\bibliographystyle{utphys}   
\bibliography{bibliography_v2} 

\newpage
\appendix

\section{Calibration of chip temperatures within the DCS software}\label{appendix:temp}
Monitoring the temperature of the ALPIDE chips is a way to spot potential issues with the cooling of individual chips. Moreover, it opens the possibility to perform correlation studies between the temperature and other observables, such as the threshold. This appendix describes in detail how the calibration of the ALPIDE temperature sensor, and the temperature measurement, are performed. 
The temperature sensor featured in the ALPIDE chips needs to be calibrated to obtain readings in degrees Celsius. This is fully performed within the DCS software. The full calibration can be summarized into two fundamental steps:
\begin{enumerate}
    \item indirect measurement of \texttt{AVDD} through the voltage DAC \texttt{VTEMP};
    \item extraction of the calibration parameters using an external temperature reference.
\end{enumerate}

\subsection{Measurement of AVDD through VTEMP} \label{sec_app:avdd_vtemp}
\begin{figure}[h!]
    \centering
    \includegraphics[width=0.5\linewidth]{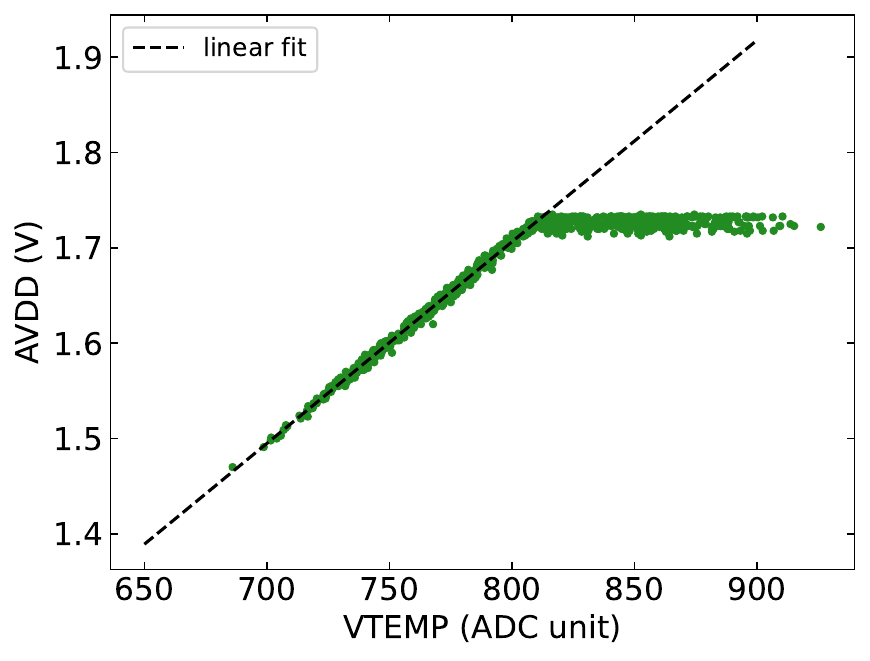}
    \caption{\texttt{AVDD} vs. \texttt{VTEMP} measurement with the internal ADC of the ALPIDE chips of an OB stave, for different voltage values set at the Power Unit. The black dashed line represents the linear fit to data for \texttt{AVDD}~$<$~1.72~V.}
    \label{fig:AVDDcalib}
\end{figure}

The \texttt{AVDD} values were measured through the voltage DAC \texttt{VTEMP} for the chips of an OB stave by setting different voltages at the PU. The \texttt{AVDD} and \texttt{VTEMP} register outputs are read out through the ADC, and the result of this measurement is shown in Fig. \ref{fig:AVDDcalib}. It is possible to see that \texttt{AVDD} depends linearly on the \texttt{VTEMP} output in ADC units with a saturation at 1.72~V above 800 ADC units. The saturation comes from the limited dynamic range of the ADC-internal DAC as explained in Sec.~\ref{sec:alpide}. 
The black dashed line represents the linear fit to data for \texttt{AVDD} $<$ 1.72 V. The line parameters are used to convert the VTEMP output from ADC units to Volts. 

\subsection{Temperature sensor calibration procedure} \label{sec_app:parameters}
A temperature reading in degrees Celsius can be obtained, for each chip, by retrieving the temperature sensor output in ADC units and the ALPIDE \texttt{AVDD} value and by applying the following equation:
\begin{equation} \label{tempss}
    \mathrm{T(^{\circ}C) = (m_{cooling} \cdot T_{ADC}(AVDD_{ref}) + q_{cooling}) + T_{heating}}
\end{equation}
with:
\begin{equation}\label{TADC}
        \mathrm{T_{ADC}(AVDD_{ref}) =  \mathrm{T_{ADC}(AVDD)} + \mathrm{m_{ADC}} \cdot(AVDD_{ref}-AVDD)}.
\end{equation}
In the following, the parameters of Eq.~\ref{tempss} and Eq.~\ref{TADC} are described. 
\begin{itemize}
    \item $\mathrm{m_{cooling}}$ and $\mathrm{q_{cooling}}$ are extracted by measuring the temperature-sensor output in ADC units as a function of the cooling water temperature. This measurement is important to have an external reference of the temperature in degrees Celsius. 
    \begin{figure}[h!]
        \centering
        \includegraphics[width=0.5\linewidth]{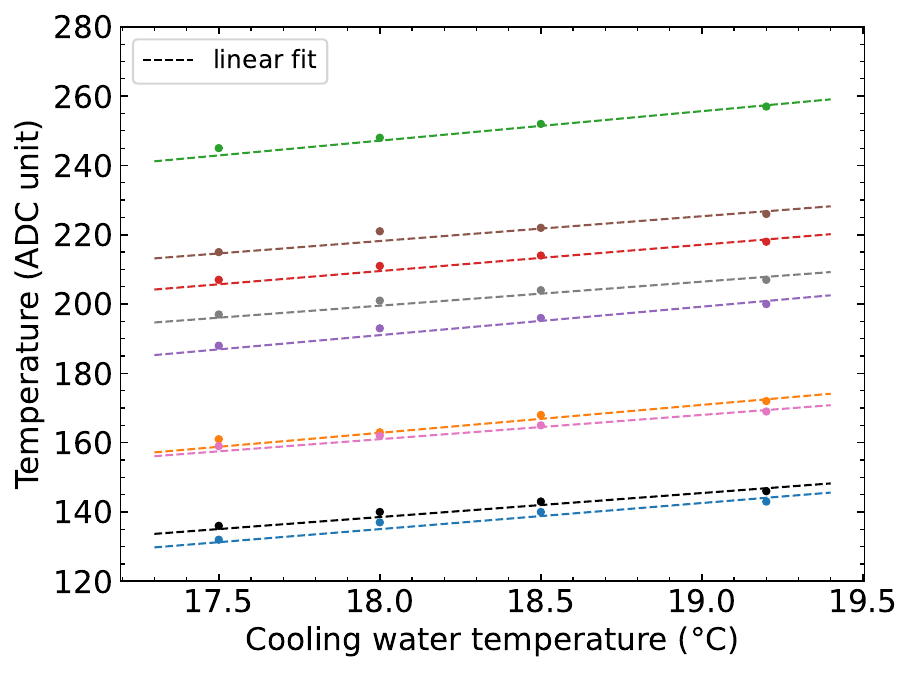}
        \caption{Temperature in ADC units vs. cooling water temperature for the 9 chips of an IB stave. Dashed lines represent linear fits performed to the data points of every chip.}
        \label{fig:temp-temp}
    \end{figure}
    Fig.~\ref{fig:temp-temp} shows how the temperature output depends linearly on the cooling water temperature. Each line corresponds to a single chip of the same stave of the IB. The \texttt{AVDD} is set to 1.8 V at the PU, and the temperature of the cooling plant is set to 17.5~°C, 18.0~°C, 18.5~°C, and 19.2~°C. The slope \textit{m} and offset \textit{q} extracted from this measurement for each chip are used to extract the parameters $\mathrm{m_{cooling}} = 1/m$ and $\mathrm{q_{cooling}} = -q/m$.

    \item $\mathrm{AVDD_{\mathrm{ref}}}$ represents the \texttt{AVDD} value at the moment of the measurement shown in Fig.~\ref{fig:temp-temp}, for which the calibration parameters $\mathrm{m_{cooling}}$ and $\mathrm{q_{cooling}}$ are applicable. The output of the temperature sensor depends on \texttt{AVDD}, hence $\mathrm{m_{cooling}}$ and $\mathrm{q_{cooling}}$ also depend on it. To ensure the applicability of Eq.~\ref{tempss} to any \texttt{AVDD} value at which the chip is operated, the output of the temperature sensor, $\mathrm{T_{ADC}(AVDD)}$, needs to be rescaled to $\mathrm{AVDD_{ref}}$ following Eq.~\ref{TADC}. The rescaled value in ADC units is $\mathrm{T_{ADC}(AVDD_{ref})}$. Fig.~\ref{fig:temp-avdd} shows the dependence of the sensor output on \texttt{AVDD} for the nine chips of an IB stave. The temperature of the cooling water is kept fixed at 18.5°C. The parameter $\mathrm{m_{ADC}}$ is the mean slope of the fit lines in Fig.~\ref{fig:temp-avdd}. 
    Means are considered in this case because the majority of the staves have chips with very similar slopes. In rare cases, this simplification can bias the mean when the performance of some chips on a stave significantly deviates from that of the other chips, as it is discussed in Sec.~\ref{sec:temp-results}. 
    \begin{figure}[h!]
        \centering
        \includegraphics[width=0.5\linewidth]{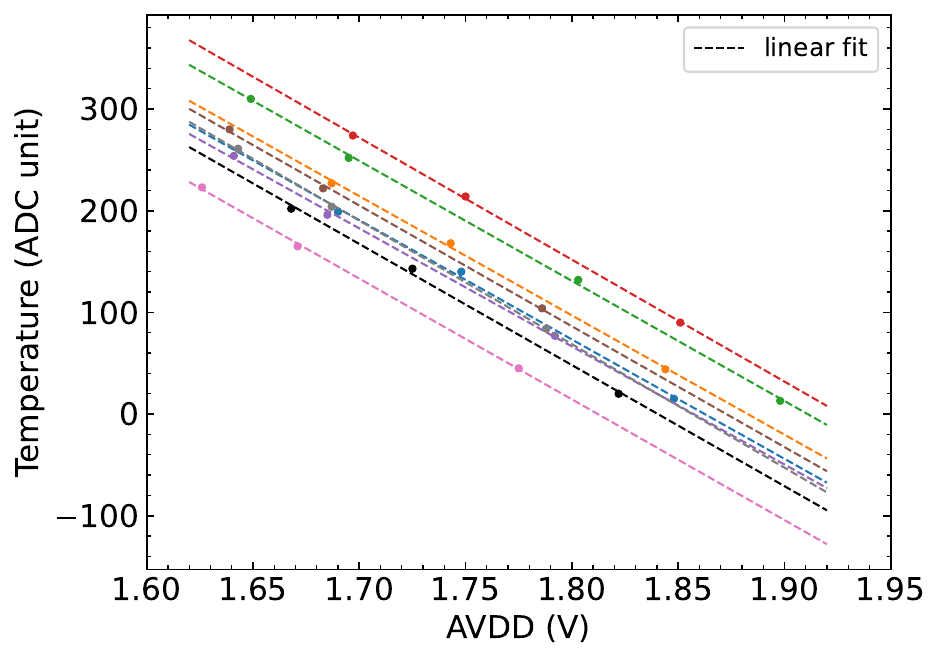}
        \caption{Measured temperature in ADC Units vs. AVDD value for the 9 chips of an IB stave. The cooling water temperature was set at 18.5 °C. The AVDD values are measured at the PU. Dashed lines represent linear fits performed to the data points of every chip. For high AVDD values, the circuitry saturates at 0 ADC units. These points were excluded from the AVDD calibration.}
        \label{fig:temp-avdd}
    \end{figure} 
    \item $\mathrm{T_{heating}}$ is the absolute correction offset, which takes into account the difference between the temperature of the cooling water and that of the chip. The assumption in this case is that the temperature of the chip is equal to the cooling water temperature when the chip is off, but is slightly higher when the chip is on. Analyzing the temperature sensor output as a function of time after stave power-on allows us to determine how the chip temperature evolves during power-up.
    Figure \ref{fig:correctionFactor} shows the rising curve of the temperature of a single chip at the power-on of the stave.
    \begin{figure}[h!]
        \centering
        \includegraphics[width=0.5\linewidth]{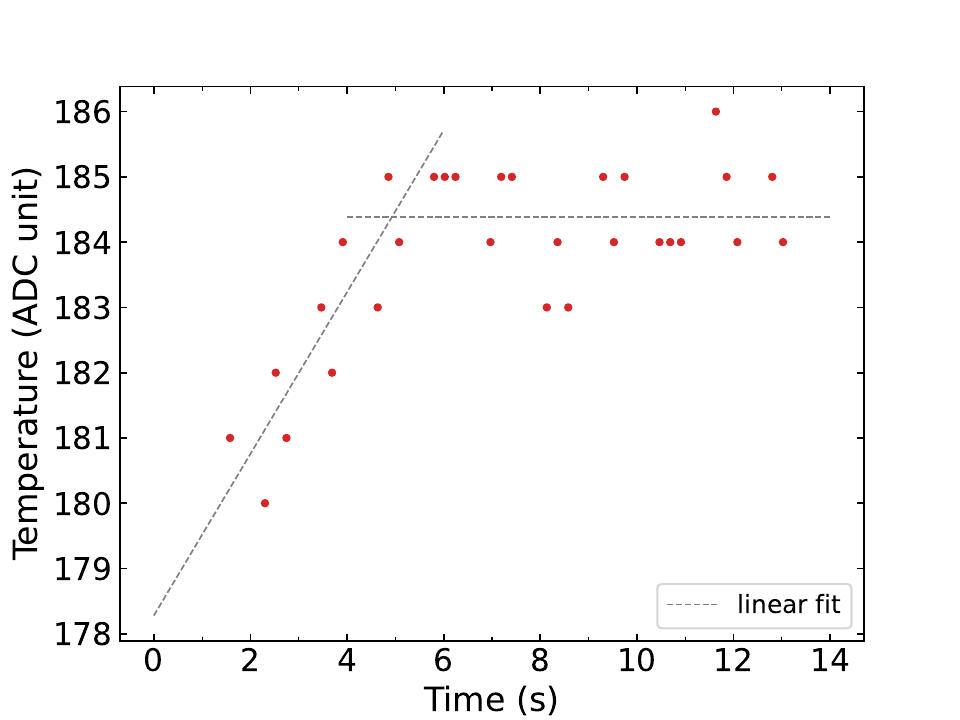}
        \caption{Temperature in ADC units vs. Time. This plot shows how the chip temperature rises at the moment of the stave power-on. Results refer to a single chip. The black lines represent the linear fits in the rising part and in the constant part of the curve.}
        \label{fig:correctionFactor}
    \end{figure}    
    By repeating the measurement on multiple chips, the offset $\mathrm{T_{heating}}$ was calculated as the mean, across chips, of the temperature difference between 10 seconds after power-on and the initial temperature at power-on (0 s):
    \begin{equation}\label{heating}
        \mathrm{ T_{heating}= \langle T(10s) - T(0s)\rangle}.
    \end{equation}
    The value of $\mathrm T_{\mathrm{heating}}$ is approximately 0.85 °C.
\end{itemize}

Due to chip-by-chip variations of the temperature measurement circuitry, a set of calibration parameters for each chip is needed. 
These parameters are stored in the ITS2 configuration database, from which they can be retrieved to obtain the ALPIDE calibrated temperature within the DCS software. They are expected to be stable over time and after irradiation.
\subsection{Temperature measurement results}\label{sec:temp-results}
\begin{figure}[h!]
    \centering
    \includegraphics[width=0.5\linewidth]{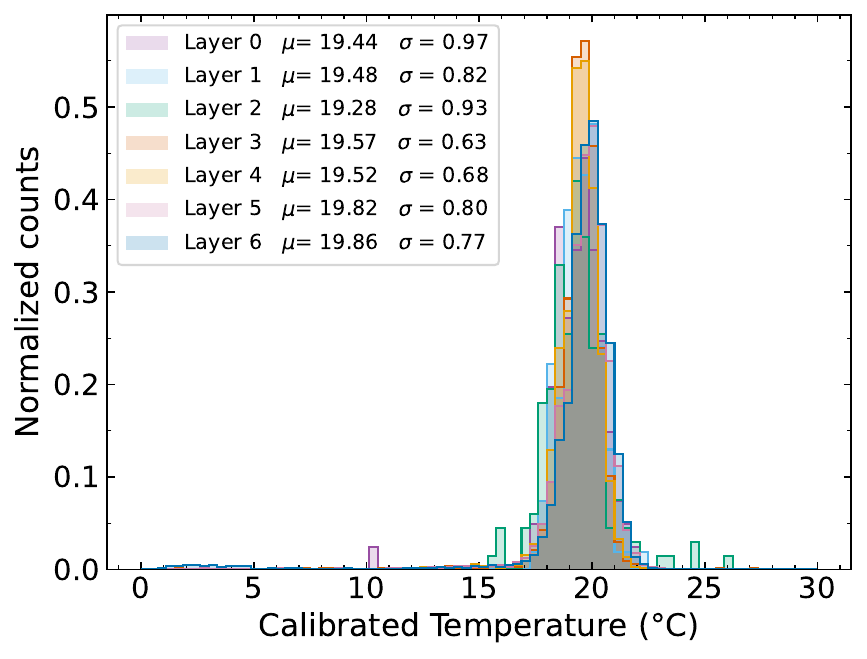}
    \caption{Temperature distribution for every ITS2 chip, layer by layer. Few outliers are due to dead/excluded chips or missing/incorrect calibration parameters. The cooling-plant water temperature set at the moment of the measurement was 18.5 °C. The mean value and standard deviation per layer as extracted from a Gaussian fit are shown in the legend. Distributions are normalised to their integral.}
    \label{fig:temperatures}
\end{figure}

Figure \ref{fig:temperatures} shows the chip temperature distribution of every ITS2 chip obtained after the temperature calibration, layer by layer.
Data were recorded after the power-on in static conditions (no data-taking ongoing), once the temperatures stabilized, with the temperature of the cooling water set at 18.5 °C. As it is possible to see, the mean temperature stays around 19--20 °C, slightly above the cooling water temperature. 
The small number of entries in the tail at lower temperature values refers to a set of chips in the detector for which the calibration did not perform well (around 2\% of the chips of the detector). This is due to the simplified calculation of $\mathrm{m_{ADC}}$ as explained in Sec. \ref{sec_app:parameters}.

%
%

\newpage
\section{The ALICE ITS Collaboration}\label{app:collab}
\begin{flushleft}
 D.~Agguiaro$^{\rm 24}$,
 G.~Aglieri Rinella\,\orcidlink{0000-0002-9611-3696}\,$^{\rm 18}$,
 L.~Aglietta\,\orcidlink{0009-0003-0763-6802}\,$^{\rm 11}$,
 M.~Agnello\,\orcidlink{0000-0002-0760-5075}\,$^{\rm 16}$,
 F.~Agnese\,\orcidlink{0000-0003-2806-6709}\,$^{\rm 42}$,
 B.~Alessandro\,\orcidlink{0000-0001-9680-4940}\,$^{\rm 27}$,
 G.~Alfarone$^{\rm 27}$,
 J.~Alme\,\orcidlink{0000-0003-0177-0536}\,$^{\rm 4}$,
 E.~Anderssen$^{\rm 32}$,
 D.~Andreou\,\orcidlink{0000-0001-6288-0558}\,$^{\rm 34}$
% ...

\bigskip

\textbf{\Large Collaboration Institutes}

\bigskip

% ...
$^{4}$ Department of Physics and Technology, University of Bergen, Bergen, Norway\\
% ...
$^{11}$ Dipartimento di Fisica dell'Universit\`{a} and Sezione INFN, Turin, Italy\\
% ...
$^{16}$ Dipartimento DISAT del Politecnico and Sezione INFN, Turin, Italy\\
% ...
$^{18}$ European Organization for Nuclear Research (CERN), Geneva, Switzerland\\
% ...
$^{24}$ INFN, Sezione di Padova, Padova, Italy\\
% ...
$^{27}$ INFN, Sezione di Torino, Turin, Italy\\
% ...
$^{32}$ Lawrence Berkeley National Laboratory, Berkeley, California, United States\\
% ...
$^{34}$ Nikhef, National institute for subatomic physics, Amsterdam, Netherlands\\
% ...
$^{42}$ Universit\'{e} de Strasbourg, CNRS, IPHC UMR 7178, F-67000 Strasbourg, France, Strasbourg, France\\
% ...
\bigskip

\end{flushleft}

\end{document}